\documentclass[aps,showpacs,twocolumn,amsmath,amssymb,eqsecnum]{revtex4}
\usepackage{graphicx}
\usepackage{epstopdf}

\begin{document}
\title{Finite-size effects on the behavior of the
susceptibility \\ in van der Waals films under  $+,+$ boundary
conditions
}
\author{Daniel Dantchev$^{1,2}$\thanks{e-mail:
daniel@imbm.bas.bg},  Joseph Rudnick$^{2}$\thanks{e-mail:
jrudnick@physics.ucla.edu}, and M. Barmatz$^{3}$\thanks{e-mail:
Martin.B.Barmatz@jpl.nasa.gov}} \affiliation{ $^1$Institute of
Mechanics - BAS, Academic Georgy Bonchev St. building 4,
1113 Sofia, Bulgaria,\\
$^2$ Department of Physics and Astronomy, UCLA, Los Angeles,
California 90095-1547, USA,\\
$^3$ Jet Propulsion Laboratory, California Institute of Technology,
Pasadena, California 91109-8099, USA}
\date{\today}

\begin{abstract}
We study critical point finite-size effects in the case of the  susceptibility of a film in which interactions are characterized by a van der Waals-type power law tail. The geometry is appropriate to a slab-like system with two bounding surfaces. Boundary conditions are consistent with surfaces that both prefer the same phase  in the low temperature, or broken symmetry, state. We take into account both interactions within the system and interactions between the constituents of the system  and the material surrounding it. Specific predictions are made with respect to the behavior of a $^3$He and $^4$He films in the vicinity
of their respective liquid-vapor critical points.

\end{abstract}

\maketitle

\section{Introduction}

Confinement to a finite volume introduces a variety of
modifications to the behavior of a system in the vicinity of a phase
transition. The existence of a single bounding surface leads to
surface phase transitions \cite{B83,D86,P2004}, critical adsorption
\cite{B83,D86}, wetting \cite{Di88} and interface phenomena
\cite{J86}. The requirement that the system is of finite extent in
one or more directions generates effects associated with finite-size
scaling theory \cite{F71,Ba83,P90,BDT2000}, including shifts in
critical points, dimensional crossover, rounding of phase
transitions and also to such phenomena as capillary condensation
\cite{capcond} and the interface delocalization phase transition
\cite{deltran,BLM2003}.

In addition, the nature of the interaction within the system and
between the system and the surrounding world influences leading and
sub-leading thermodynamic behavior at and near a critical point
\cite{DR2001,D2001,CD2002,DKD2003,T2005,DDG2006}.

In this article we will discuss finite-size effects as
they apply to the susceptibility of a film  of a non-polar
fluid having a thickness $L$ in which the intrinsic interaction
$J^l$ is of the van der Waals type, decaying with distance
$r$ between the molecules of the fluid as $J^l \sim r^{-(d+\sigma)}$.
Here $d$ is the dimensionality of the system while $\sigma>2$
is a parameter characterizing the decay of the interaction.
The film is bounded by a substrate that interacts with the
fluid with a similar van der Waals type forces, i.e. of
the type $J^{l,s}\sim z^{-\sigma_s}$, where $z$ is
the distance from the boundary of the system while
$\sigma_s>2$ characterizes the decay of the
fluid-substrate potential. For realistic
fluids $d=\sigma=\sigma_s=3$. The discussion in this
paper will be quite general, but we will be principally
interested in an Ising type model, which is commonly
utilized to represent a non-polar fluid.

As a specific application, we will  estimate the parameters required
for the prediction of the behavior of the finite-size susceptibility
in the case of $^3$He and $^4$He films near their respective
liquid-vapor critical points. In line with an experimental
investigation of the phenomena that we discuss here, we assume that
the film is surrounded by gold surfaces. We will take into account
both the van der Waals type interaction between the atoms of the
$^3$He (or $^4$He) and the corresponding interaction between the Au
atoms and $^3$He (or $^4$He) atoms.

According to general scaling arguments \cite{D86}, \cite{PL83},
the finite-size behavior of the susceptibility in a film of a fluid
subject to  $(+,+)$ boundary conditions is
\begin{figure}[h]
\includegraphics[width=\columnwidth]{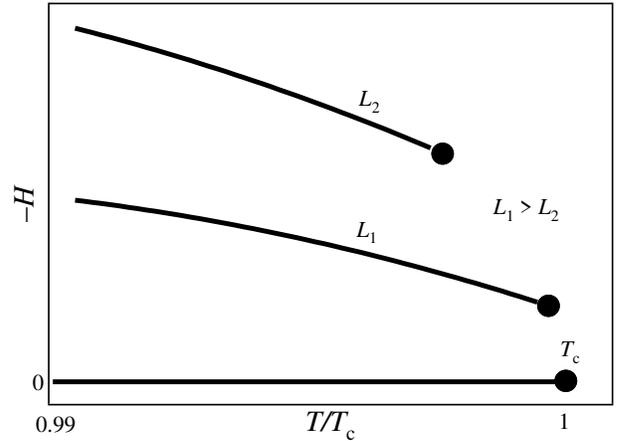}
\caption{A schematic ($H$,$T$) phase diagram near the critical point
of a $d$-dimensional system with $(+,+)$ boundary conditions
\cite{BLM2003,BL91}. The line $H=0$, $T<T_c$ represents the bulk
phase coexistence line. (In the case of a fluid system the vertical
-$H$ axis then corresponds to the vapor side of the fluid-vapor
coexistence curve.) The upper phase coexistence line is for a system
with thickness $L_2$, while the lower one is for a system with
thickness $L_1$, where $L_1>L_2$. Away from the critical region the
shift in the phase boundary is proportional to $L^{-1}$, while
within it is proportional to $L^{-\Delta/\nu}$. As is shown, the
lines of a first-order phase transition (with respect to the
external bulk magnetic field) end at critical points of the
$(d-1)$-dimensional system, the positions of which depend on $L$, on
the amplitude of the (long-range) surface field and on the presence
or absence of long-range interactions between the spins of the
system. \label{pd}}
\end{figure}
\begin{widetext}
\begin{eqnarray} \label{hypoffss}
\chi(t,h,L) &-& \chi_{\rm bulk}(t,h)-L^{-1}\left[\chi_{{\rm
surface},1}(t,h)+\chi_{{\rm surface},2}(t,h)\right]= \\
&& L^{\gamma/\nu}X(L/\xi_t,a_h h L^{\Delta/\nu}, b
L^{2-\sigma-\eta},h_{s} L^{(d+2-\eta)/2-\sigma_s};a_\omega
L^{-\omega}), \nonumber
\end{eqnarray}
\end{widetext}
where $\chi_{\rm bulk}(t,h)$ is the bulk susceptibility, $\chi_{{\rm
surface},1}(t,h)$ and $\chi_{{\rm surface},2}(t,h)$ are the surface
susceptibilities, $\xi_t(T)=\xi_\infty(T \rightarrow
T_c^+,h=0)\simeq \xi_0^+|t|^{-\nu}$ is the bulk correlation length,
$t=(T-T_c)/T_c$ is the reduced temperature, $T_c$ is the bulk
critical temperature, $\xi_0^+$, $a_h$, $b$, $h_{w,s}$ and
$a_\omega$ are nonuniversal metric constants, while $\nu$, $\Delta$,
$\eta$ and $\omega$ are the universal critical exponents for the
corresponding short-range system. For a schematic phase diagram in the ($H$,$T$) plane of
the system see Fig. \ref{pd}. Assuming $\sigma>2-\eta$ and
$\sigma_s>(d+2-\eta)/2$, one can expand (\ref{hypoffss}), which leads
to
\begin{widetext}
\begin{eqnarray}\label{chiexpanded}
\chi(t,h,L)& \simeq & \chi_{\rm bulk}(t,h)
+L^{-1}\left[\chi_{{\rm surface},1}(t,h)+\chi_{{\rm surface},2}(t,h)\right]\\
&& +L^{\gamma/\nu}\left\{X^{\rm sr}(a_ttL^{1/\nu},a_h h
L^{\Delta/\nu};a_\omega L^{-\omega})+h_{s} L^{(d+2-\eta)/2-\sigma}
X^{\rm lr}_{s,1}(a_ttL^{1/\nu},a_h h L^{\Delta/\nu}) \right.
\nonumber
\\
& & \left. +L^{-(\sigma-2+\eta)} \left[h_{s}^2 L^{-(\sigma - d)}
X^{\rm lr}_{s,2}(a_ttL^{1/\nu},a_h h L^{\Delta/\nu})+b X^{\rm
lr}_b(a_ttL^{1/\nu},a_h h L^{\Delta/\nu})\right]\right\}, \nonumber
\end{eqnarray}
\end{widetext}
where $X^{\rm sr}$ is the (universal) scaling function
characterizing the truly short-range system, while the remaining
part in (\ref{chiexpanded}) describes the contributions due to the
(subleading) long-range part of the interaction. While it is well established that $X^{\rm sr}$ tends to zero as
$\exp[-aL/\xi_\infty]$ with $a$ a constant when $L/\xi_\infty \gg
1$, the functions $X^{\rm lr}_{s,1}$, $X^{\rm lr}_{s,2}$ and $X^{\rm
lr}_b$ are expected to decay only in a power-law-in-$L$ way.

 Because of that, whenever
$L/\xi_\infty\gg 1$ the functions $X^{\rm lr}_{s,1}$, $X^{\rm
lr}_{s,2}$ and $X^{\rm lr}_b$ will determine the leading-in-$L$
finite-size contributions to the susceptibility of the system, and,
therefore, the {\it leading} finite-size behavior of the
susceptibility. Note that due to the lack of ``+''$\Leftrightarrow
$``-'' symmetry in the surface field given  $(+,+)$ boundary conditions,
one finds in the
susceptibility a term {\it linearly} proportional to $h_{w,s}$
(in addition to the term proportional to $h_{w,s}^2$).

In the case of the ``genuine'' van der Waals interaction
$d=\sigma=\sigma_s=3$. For the three-dimensional Ising
model~\cite{JJ2002} $\eta= 0.034$, $\gamma=1.2385$, $\nu=0.631$,
$\alpha=0.103$, $\beta=0.329$, $\theta \equiv \omega \nu =0.53$.

In the remainder of this article we will consider only the case of
identical substrates bounding the system. That is, we assume
$\chi_{{\rm surface},1}(t,h)=\chi_{{\rm surface},2}(t,h) \equiv
\chi_{{\rm surface}}(t,h)$. Furthermore, for the bulk susceptibility
one has \cite{KR84}
\begin{equation}\label{chi_bulk}
\chi_{\rm bulk}(t,h)=\chi_{\rm bulk, \, sr}(t,h)+\chi_{\rm bulk, \,
lr}(t,h),
\end{equation}
where \cite{KR84,KG81,ZBH2003}
\begin{eqnarray}\label{chi_bulk_sr}
&& \chi_{\rm bulk, \, sr}(t\to 0,h=0)=\Gamma^\pm
|t|^{-\gamma}\left[1+ \right. \\
&& \left. \Gamma^\pm_1 |t|^\theta + \Gamma^\pm_2 |t|^{\theta_2}
+\Gamma^\pm_{1,2} |t|^{2\theta}+ \Gamma^\pm_{\gamma,\alpha}
|t|^{\gamma-\alpha}+\cdots\right], \nonumber
\end{eqnarray}
and
\begin{equation}\label{chi_bulk_lr}
\chi_{\rm bulk, \, lr}(t\to 0,h=0)=\Gamma^\pm |t|^{-\gamma}\left[
\Gamma^\pm_{\rm lr} |t|^{\theta_\sigma} (\ln |t|)^{-\sigma}\right].
\end{equation}
Here $\theta_2\simeq 0.90$ \cite{S82} is the second Wegner exponent,
while $\theta_\sigma\equiv\nu(\sigma-2+\eta)\simeq 0.65$ is the
corresponding exponent due to the long-range tail of the van der
Waals interaction. Note that $\theta_{\sigma}$
is only slightly larger than the dominant Wegner term for the Ising universality class . Apart from
the logarithmic corrections (of the predicted type $\sim(\ln
|t|)^\sigma$) the existence of such a correction to scaling bulk
term has been confirmed within the exactly solvable spherical model
\cite{DR2001}. The logarithmic term ($\sim \ln|t|$) has
been observed, however, only for $d+\sigma=6$. In \cite{DDG2006} it
has been demonstrated that such a term is a peculiarity of the model
which is due to the degeneracy, within this model, of the exponents
$\omega=4-d$ and $\omega_\sigma\equiv \theta_\sigma/\nu=\sigma-2$
(when $d+\sigma=6$). Thus, for an Ising like system with subleading
van der Waals type long-range interactions we expect that
\begin{equation}\label{chi_bulk_lr_correct}
\chi_{\rm bulk, \, lr}(t\to 0,h=0)=\Gamma^\pm |t|^{-\gamma}\left[
\Gamma^\pm_{\rm lr} |t|^{\theta_\sigma}  \right].
\end{equation}
As far as we are aware, no experimental verification of such
type bulk correction to scaling has been reported in the available
literature.

In the case of $(+,+)$ boundary conditions the surface
susceptibility is controlled by the extraordinary (or normal)
surface universality class. One has
\begin{equation}\label{chi_surface}
\chi_{{\rm surface}}(t\to 0,h=0)=\Gamma_{\rm surface}^\pm \;
|t|^{-\gamma_s},
\end{equation}
with $\gamma_s=\gamma+\nu$ \cite{B83}. Thus, for the
three-dimensional Ising universality class $\gamma_s \simeq 1.87$.
We are not aware of any study of explicit corrections to the
behavior of this quantity due to van der Waals forces.
It is clear that these forces do not suffice to change the
surface universality class. That is, the critical exponents will
remain the same as in the case of completely short-range
interactions \cite{PL83,D86}.

Let us now investigate in more detail the conditions under which the
expansion in Eq.
(\ref{hypoffss}), leading to (\ref{chiexpanded}), can be justified.
Some requirements, such as $2-\eta-\sigma<0$, $(d+2-\eta)/2-\sigma<0$,
are obvious and normally are satisfied in any realistic system for
which $d=\sigma=3$ and $\eta\ll 1$ (i.e., for the 3d Ising model in which
$\eta\simeq 0.034$ \cite{JJ2002}). Important additional conditions
arise, however, from the fact that we consider a {\it finite}
system in which power law {\it long-range} surface fields (i.e.
substrate-fluid potentials) act. The influence of those long-range surface fields
is felt everywhere in the finite system, the amplitude of the surface field being
minimum
at the center of the system. One can think of  this smallest value as a type of a
bulk field $h$, which has the effect of displacing the system from the position on
the phase diagram on which its bulk field would otherwise place it.
Taking into account contributions from both surfaces we
obtain for the contribution of the long range surface field to an effective bulk
symmetry-breaking field $h_{b,s}=2 h_{w,s} \left[L/(2\xi_0^+)\right]^{-\sigma_s}$.
Since
the bulk magnetic field scales as $h [L/\xi_0^+]^{\Delta/\nu}$ one
arrives at the  criterion that in a film the finite-size
contributions due to the long-range surface fields will be
negligible in the critical region if
\begin{equation}\label{condition1}
2 |h_{w,s}| \left[L/(2\xi_0^+)\right]^{-\sigma}
\left[L/\xi_0^+\right]^{\Delta/\nu} \ll 1, \end{equation} i.e.
\begin{equation}
\label{condition2} 2^{\sigma+1} |h_{w,s}|
\left[L/\xi_0^+\right]^{\Delta/\nu-\sigma} \ll 1.
\end{equation}
Note that $h_{w,s}>0$ corresponds to attractive walls, i.e. walls
preferring the liquid phase of the fluid  while $h_{w,s}<0$
corresponds to repulsive walls, i.e. to walls preferring the gas
phase of the fluid. More detailed discussion on that point is
presented in Appendix \ref{helium}, where we identify $h_{w,s}$ in
the framework of a mean-field type model. Using the relations
between critical exponents it is easy to show that
$\Delta/\nu=(d+2-\eta)/2$. Thus the relation (\ref{condition2}) is
consistent with the form (\ref{hypoffss}).  On the other hand,
$\Delta/\nu=d-\beta/\nu$ and, therefore,
$\Delta/\nu-\sigma=d-\sigma-\beta/\nu$. By taking into account
that for realistic systems $d=\sigma$, the condition
(\ref{condition2}) becomes
\begin{equation}\label{condition3}
2^{\sigma+1} |h_{w,s}| \left[L/\xi_0^+\right]^{-\beta/\nu} \ll 1.
\end{equation}
For most systems $\xi_0^+$ is of the order
of 3 {\AA}. Taking the values of $\beta$ and $\nu$ to be appropriate to the
3d Ising model, we obtain in the case $\sigma=3$
\begin{equation}\label{cL}
L\gg L_{\rm crit} \equiv \xi_0^+ \left(
2^{\sigma+1}|h_{w,s}|\right)^{\nu/\beta} \simeq 612 \,
|h_{w,s}|^{1.918}\; \mbox{\AA}.
\end{equation}
Later on in this article we will discuss the determination of  the magnitude
of $|h_{w,s}|$ for the cases
of $^3$He and $^4$He films bounded by Au surfaces. For the moment
we note that in such systems $|h_{w,s}| \simeq 4$ (see Appendix
\ref{helium}). The condition $L\gg 9000$\;{\AA} must be met
if finite-size effects due to the van der Waals interactions are
to be neglible within the critical region of an $^3$He or
$^4$He film.
If  $L\sim 15000$\;{\AA}, which is experimentally realizable, then
one expects van der Waals finite-size effects to play an
essential role everywhere within the critical region. This
implies that the value of the finite-size susceptibility at $T_c$
will depend on $L$ and on the choice of the bounding substrate (i.e.
on the value of $h_{w,s}$).

Away from $T_c$ one expects
\begin{equation}\label{decompositiondelta}
    \chi=\chi_{{\rm bulk}}+\frac{1}{L}\chi_{{\rm
    surface}}+\frac{1}{L^\sigma}\chi_{\rm Hamaker} \cdots
\end{equation}
In a fluid system below $T_c$ the Hamaker term comprises
contributions due to both the fluid-fluid and the substrate-fluid
interaction (the substrate-substrate interaction does not contribute
because the substrate density $\rho_s$ does not depend on the
magnetic field). 
In a fluid system the density is the quantity that couples via the
van der Waals type interaction. The average local density
is non-zero both below and above the critical point, whereas in a
magnetic system $m_0(T>T_c,H=0)=0$. Because of this, there exists a
difference in the finite-size behavior of the  fluid and the
magnetic systems when $T>T_c$. One can argue that for
a magnetic system in which $T>T_c$ and $h=0$ \cite{MDB2004}
\begin{equation}\label{decompositiondeltamagnetic}
    \chi=\chi_{{\rm bulk}}+\frac{1}{L}\chi_{{\rm
    surface}}+\frac{1}{L^{\sigma +1}} \tilde{\chi}+
     \cdots
\end{equation}
Below $T_c$ Eq. (\ref{decompositiondelta}) is again valid. As we
see, the terms of the order of $L^{-(d+\sigma)}$ will never be
important if substrate potentials are present in the system. Such
terms, however, give rise to the leading finite-size contributions away
from the critical region in the case of periodic boundary conditions
\cite{DR2001,D2001,CD2002,DDG2006}. Contributions from the
fluid-fluid potential are proportional to $m^2$ and, therefore, will
be of importance only {\it below} the critical temperature (if $|H|
\ll 1$). They then constitute one of the Hamaker terms.

Expressions similar to Eq. (\ref{chiexpanded}) have been shown to
describe the finite-size behavior of the susceptibility in a fully
finite system subject to periodic boundary conditions---in which
case $h_{w,s}\equiv 0$---both in the instance of the exactly
solvable spherical model \cite{DR2001,D2001,DDG2006} and  via
$\varepsilon$-expansion techniques (up to first order in
$\varepsilon$), in $O(n)$ models \cite{CD2002}.

\begin{figure}[htb]
\includegraphics[width=\columnwidth]{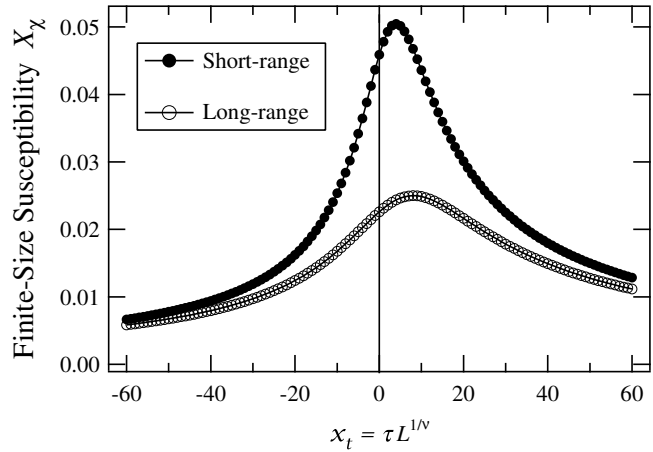}
\caption{A comparison between the behavior of the scaling functions
of the susceptibility $X_\chi(x_t)$ for a fluid in which both the
substrate-fluid and the fluid-fluid interactions are {\it
short-range} and the same quantity for a fluid in which both
interactions are {\it long-range}. The thickness of the film is $L
a$, where $L=3000$, while $a$ is the average distance between the
molecules of the fluid. The scaling variable is $x_t=\tau L^{1/\nu}$, where $\tau=1 - T_c/T$ with $T_c$ being the critical temperature of the
corresponding short- or long-range system. The curves
differ essentially from each other, in that they have different
magnitudes at $T_c$. The maximum of the scaled susceptibility is
also shifted.
  \label{3000_sr_lr_comp.eps}}
\end{figure}

Figure \ref{3000_sr_lr_comp.eps} displays the behavior of the
susceptibility in a film in which the interactions are completely
short-range and of a film in which both the fluid-fluid and the
fluid-substrate interactions are long-range, the latter case
corresponding to the actual experimental situation. We observe that
the curves behave quantitatively differently in the two systems. At
$T_c$ and at coexistence, the susceptibility of the van der Waals
type system  is severely suppressed in comparison to that of the
short-range system. As explained above and as we will see in more
detail below, the magnitude of the maximum depends on the strength
of the substrate-fluid coupling and is, therefore, {\it not}
universal. Furthermore, there is a shift in the position of the
maximum of the susceptibility. We expect that the curves shown here
will resemble those obtained via experimental investigations.

In this article, we will explore the mechanisms underlying the
similarity of the susceptibility in a fluid film near its bulk
critical point to the susceptibilities plotted in Fig.
\ref{3000_sr_lr_comp.eps}. We investigate the behavior of the
susceptibility as a function of the temperature, the external bulk
magnetic field, the thickness of the film and the strength of the
fluid-substrate interaction.

The structure of the article is as follows. First, in Section
\ref{model} we present a precise formulation of the model of
interest along with the analytical expressions needed for its
numerical treatment. The principal results for the behavior of the
finite-size susceptibility are presented in Section \ref{results}. Some technical
details---including sums containing van der Waals type
potentials---can be found in Appendix \ref{profiles}.
An estimation of the parameters needed in order to apply the current
results for the behavior of the susceptibility in $^3$He and $^4$He
films is contained in Appendix \ref{helium}.

\section{The model}
\label{model}

Perhaps the simplest model that captures the basic features of
systems with van der Waals interactions  is a modification of the
one utilized by Fisher and Nakanishi in their mean-field
investigation of short-range systems \cite{NF82,NF83}. One starts
with the following form of the functional for the free energy of the
lattice system
\begin{widetext}
\begin{equation}\label{freeenergyfunctionalstarting}
\beta F=\sum_{{\bf r}}\left\{\frac{1+m({\bf r})}{2}
\ln\left[\frac{1+m({\bf r})}{2}\right]+\frac{1-m({\bf r})}{2}
\ln\left[\frac{1-m({\bf r})}{2}\right]\right\}
 -\sum_{{\bf r}}h({\bf r}) m({\bf
r})-\frac{1}{2}\sum_{{\bf r}, {\bf r}'}K({\bf r},{\bf r}')m({\bf
r})m({\bf r}'),
\end{equation}
\end{widetext}
where  $K({\bf r},{\bf r}')=\beta J^l({\bf r},{\bf r}')$ is
the non-local coupling between magnetic degrees of freedom,
$h({\bf r})$ is an external magnetic field and the
magnetization $m({\bf r})$ is to be treated as a variational parameter.
Note that in Eq.\,(\ref{freeenergyfunctionalstarting}) the term in curly
brackets corresponds to the entropic contributions, while the other
terms are directly related to the interactions present in the system.

The variation of (\ref{freeenergyfunctionalstarting})  with respect
to $m({\bf r})$ leads to the following equation of state for our system
\begin{equation}\label{eqofstate}
    m^*({\bf r})=\tanh \left[\sum_{{\bf r}'}K({\bf r},
    {\bf r}')m^*({\bf r}')+h({\bf r})
    \right].
\end{equation}
This equation can be solved numerically by applying the
Newton-Kantorovich method. One is able to treat reasonably thick
films, with $L/a$ of the order of $3000$ layers, corresponding to
the experimental setup that we envisage as an example in our study.
Its
solution for a given geometry and external fields $h({\bf r})$
determine the order-parameter profile $m^*({\bf r})$ in the system.

We will be interested in a system with a {\it film} geometry.
Then if ${\bf r}=\{{\bf r}_\parallel,z\}$ and $h({\bf r})\equiv
h({\bf r}_\|,z)=h(z)$ one has, because of the symmetry of the
system, $m({\bf r})\equiv m({\bf r}_\|,z)=m(z)$. The magnetization
profile now depends only on the coordinate perpendicular to the plates
bounding the van der Waals system. In this case, Eq. (\ref{eqofstate}),
which describes the behavior of the magnetization profile, becomes
\begin{equation}\label{mpfilm}
m^*(z)=\tanh\left[\beta \sum_{{\bf r}'}J({\bf r}-{\bf
r}')m^*(z')+h(z)\right],
\end{equation}
where ${\bf r}'=({\bf r}_\|',z')$. Obviously, the above equation is equivalent to
\begin{equation}\label{theeqwithG}
  m^*(z)=\tanh\left[\beta \sum_{z'=0}^L \hat{{\cal J}}(z-z') m^*(z')+h(z)\right],
\end{equation}
where
\begin{equation}\label{Gdef}
\hat{{\cal J}}(z)\equiv \sum_{{\bf r}_\|'}J({\bf r}_\|-{\bf
r}_\|',z)=\sum_{{\bf r}_\|}J({\bf r}_\|,z).
\end{equation}

We will now assume that the fluid molecules occupy the region in space
characterized by $0\le z \le L$ and that the layers $z=0$ and $z=L$
satisfy the $(+,+)$ boundary conditions, i.e. $m(0)=m(L)\equiv 1$.
The number of layers containing spins that can fluctuate is, therefore, $L-1$.

Equation (\ref{theeqwithG}) is equivalent to
\begin{equation}\label{todiff}
{\rm arctanh} [m^*(z)]=h(z)+\beta \sum_{z'}\hat{{\cal J}}(z-z')
m^*(z').
\end{equation}
Taking the functional derivative of the both sides of Eq. (\ref{todiff})
with respect to the field $h(z^*)$ applied to the layer $z^*$, we obtain
\begin{equation}\label{resfunction}
\frac{G(z,z^*)}{1-m^2(z)}=\delta_{z,z^*}+ \beta
\sum_{z'}\hat{{\cal J}}(z-z')G(z',z^*),
\end{equation}
where
\begin{eqnarray}\label{Grdef}
G(z,z^*)&\equiv& \frac{\delta m(z)}{\delta h(z^*)}\\
&=&\sum_{{\bf r}_\|^*}\langle S({\bf 0},z)S({\bf
r}_\|^*,z^*)\rangle-\langle S({\bf 0},z)\rangle \langle S({\bf
r}_\|^*,z^*)\rangle.\nonumber
\end{eqnarray}
Equation (\ref{resfunction}) can be rewritten in the form
\begin{equation}\label{eq_resp_function}
\sum_{z'} \left[\frac{\delta_{z,z'}}{1-m^2(z')}-\beta \hat{{\cal
J}}(z-z')\right]G(z',z^*)=\delta_{z,z^*}.
\end{equation}
Now it is clear that the solution of the above equation with respect
to $G(z',z^*)$ is
\begin{equation}\label{sol_resp_function}
G(z',z^*)=\left(\bf{R}^{-1}\right)_{z',z^{*}},
\end{equation}
where $\bf{R}^{-1}$ is the inverse matrix of the matrix $\bf{R}$
with elements
\begin{equation}\label{R_matrix_def}
    R_{z,z'}=\frac{\delta_{z,z'}}{1-m^2(z')}-\beta \hat{{\cal
J}}(z-z').
\end{equation}

One can define the ``local'' susceptibility
 \begin{eqnarray}\label{chideflocal}
 \chi(z) & \equiv & \sum_{z^*} G(z,z^*) \nonumber \\ & = & \sum_{{\bf
r}^*}\langle S({\bf 0},z)S({\bf r}_\|^*,z^*)\rangle-\langle S({\bf
0},z)\rangle \langle S({\bf r}_\|^*,z^*)\rangle. \nonumber \\
\end{eqnarray}
Obviously, $\chi \equiv \sum_z\chi(z)/(L+1)$ is the total susceptibility
of the system per unit spin (see Eq.
(\ref{chideflocal})). Thus, one has
\begin{equation}\label{chi_sol}
\chi(z)=\sum_{z^*}\left(\bf{R}^{-1}\right)_{z,z^{*}},
\end{equation}
and
\begin{equation}\chi
=\frac{1}{L+1}\sum_{z,z^*}\left(\bf{R}^{-1}\right)_{z,z^{*}}.
\end{equation}

In Appendix \ref{profiles} we demonstrate that the function $\hat{{\cal J}}(z)$
can be written in the form
\begin{eqnarray}\label{J}
\hat{{\cal J}}(z) &=&J^l \left[c_{d-1}\delta(z)+c_{d-1}^{\rm
nn}\left[\delta(z-1)+\delta(z+1)\right] \right. \nonumber \\
&&\left. +{\cal G}_d(z) \theta(z-2)\right],
\end{eqnarray}
where $\delta(z)$ is the discrete delta function, while $\theta(z)$ is the
Heaviside function. Explicitly for $d=\sigma=3$  one has (see Appendix \ref{profiles})
\begin{equation}\label{c2t}
c_2=\sum_{{\bf n}\in {\mathbb Z}^2}\frac{1}{1+|{\bf n}|^6}\simeq
3.602,
\end{equation}
\begin{widetext}
\begin{eqnarray}\label{c2nnt}
c_2^{nn}&=&-\frac{8}{3} \pi
\left[(-1)^{1/3}K_0(\sqrt{2-2i\sqrt{3}\pi})-(-1)^{2/3}K_0(\sqrt{2+2i\sqrt{3}\pi})\right]
+ \frac{\pi}{3}\left(\frac{\pi}{\sqrt{3}}-\ln 2\right) \approx
1.183,
\end{eqnarray}
and
\begin{equation}
\label{gt} {\cal G}_3(x)=\frac{\pi}{3}\left[\sqrt{3}\ \arctan{
\frac{\sqrt{3}}{2x^2-1}}-\ln\left(1+\frac{1}{x^2}\right)+
\frac{1}{2}\ln\left(1-\frac{1}{x^2}+\frac{1}{x^4}\right)\right].
\end{equation}
\end{widetext}
The layer magnetic field $h(z)$ is the only quantity in Eq. (\ref{todiff}) the exact form of which
still requires specification. We
take it to be of the form
\begin{equation}\label{hz}
h(z)=\frac{h_{\rm w,s}}{(z+1)^3}+\frac{h_{\rm w,s}}{(L+1-z)^3}, \ \
1\le z \le L-1,
\end{equation}
where $h_{\rm w,s}$ reflects the relative strength of the fluid-wall
and fluid-fluid interactions (see Eq. (\ref{sdef}) below). The above
expression is consistent with the fact that the substrate occupies
the region $\mathbb{R}^{d-1}\times [L+1,\infty] \; \cup \;
\mathbb{R}^{d-1}\times [-(L+1),-\infty]$. For $^3$He and $^4$He
bounded by Au surfaces we show in Appendix \ref{helium} that $h_{\rm
ws}=4$.

\section{Numerical results and finite-size scaling analysis}
\label{results}

To determine the total susceptibility $\chi(T,H|L,h_{w,s})$ and its
"scaling function"
\begin{equation}\label{chi_def}
X_\chi\equiv L^{-\gamma/\nu} \chi(T,H|L,h_{w,s})
\end{equation}
in a fluid film with thickness $L$, one has to solve Eq.
(\ref{todiff}) for $1 \le z \le L-1$, which allows one to construct
the matrix $\bf{R}$ with the use of Eq. (\ref{R_matrix_def}).
Finally, on the basis of Eq. (\ref{chi_sol}) one obtains the local
susceptibility $\chi(z)$, $1\le z\le L-1$, and, summing, the total
susceptibility. We recall that within the mean field treatment one
has $\beta=\nu=1/2$, $\gamma=1$, $\eta=0$ and, for the extraordinary
surface transition, $\gamma_s=3/2$ \cite{B83}.

The analytic solution to the set of coupled nonlinear equations for
the magnetization profile is, at present,  known only for purely short-range,
continuous, systems \cite{M97}. We review this solution in section \ref{psrs}.
Even in this case the the finite-size susceptibility has to be determined
numerically. The results are, again,  presented in section \ref{psrs}.
Numerical methods appear to be unavoidable in order to solve the equations for the magnetization profile in the
case of the long-range interactions of the van der Waals type. The results in this case will be discussed in section \ref{slri}.

\subsection{The model with purely short-range interactions for $H=0$}
\label{psrs}

The equations to be solved in the continuum version of the purely short-range model
of a mean-field
Ising strip under $(+,+)$ boundary conditions are
\begin{eqnarray}
  1 &=& \left[1-\frac{\beta}{\beta_c} -
  \beta \frac{\partial^2}{\partial z^2}
  +m^2(z)\right]\chi(z), \\
  0&=& \left[1-\frac{\beta}{\beta_c} -
  \beta \frac{\partial^2}{\partial z^2}
  +\frac{1}{3} m^2(z)\right]m(z).
\end{eqnarray}
Because conditions are identical at both bounding surfaces of the
system, the solutions of the above equation satisfy  $m'(L/2)=0$ and
$\chi'(L/2)=0$. The magnetization profile is known exactly
\cite{M97}: \begin{enumerate}

\item[a)] when $tL^2\ge \pi^2$ [with $t=(T-T_c)/T_c$]
\begin{equation}\label{mz}
m(z)=\frac{2 K(k)}{L}\frac{{\rm dn}(\zeta;k)}{{\rm sn}(\zeta;k)},
\end{equation}
where
\begin{equation}\label{tk}
    tL^2=[2K(k)]^2(2k^2-1), \qquad \zeta=[2K(k)/L]z,
\end{equation}
and $k^2 \ge 0$.

\item[b)] when $tL^2 \le \pi^2$
\begin{equation}\label{mzd}
m(z)=\frac{2 K(k)}{L}\frac{1}{{\rm sn}(\zeta;k)},
\end{equation}
where
\begin{equation}\label{tkb}
    tL^2=-[2K(k)]^2(k^2+1), \qquad \zeta=[2K(k)/L]z,
\end{equation}
and $k^2 \ge 0$.
\end{enumerate}
Here $K(k)$ is the complete elliptic integral of the first kind,
${\rm dn}(\zeta;k)$ and ${\rm sn}(\zeta;k)$ are the Jacobian
$\Delta$ amplitude and the sine amplitude functions, respectively.
The bulk critical point $T=T_c$ corresponds to $k^2=1/2$.
The above expressions are consistent with the following scaling
form for the order parameter:
\begin{equation}\label{sf}
m(z)=L^{-\beta/\nu}X_m(\frac{z}{L},tL^{1/\nu}),
\end{equation}
with $\beta=\nu=1/2$.

\begin{figure}[htb]
\includegraphics[width=\columnwidth]{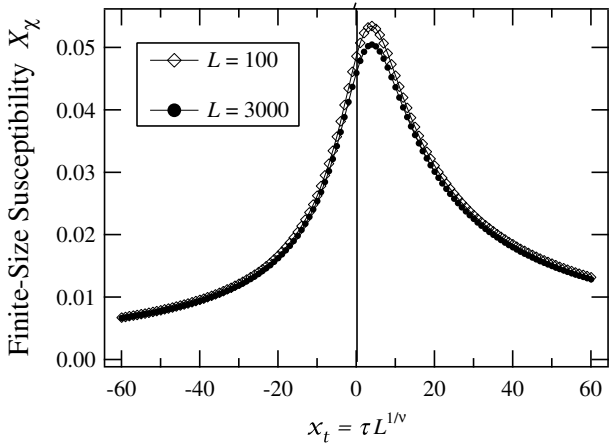}
\caption{The upper dotted line, for $L=100$, and the lower dotted line, for $L=3000$,
illustrate the behavior of the scaling function of the susceptibility $X_\chi(x_t)$
for a fluid in which both the substrate-fluid and the fluid-fluid interactions
are {\it short-range}. The scaling variable
is $x_t=\left(1-T_c/T \right)L^{1/\nu}$.
Note that the maximum of the finite-size curves is at
$T>T_c$ which is due to the stabilizing effect of the
$(+,+)$ boundary conditions on the order parameter in
the finite system. We observe that the both curves
differ at most around their maximal value. As it will
be demonstrated in Fig.
{\ref{100_3000_xi_sr_comp_norm_best}} the principal
reason for this deviation is the improper choice of
the value of the distance $L$ between the plates.
  \label{100_3000_xi_sr_comp_best}}
\end{figure}

\begin{figure}[htb]
\includegraphics[width=\columnwidth]{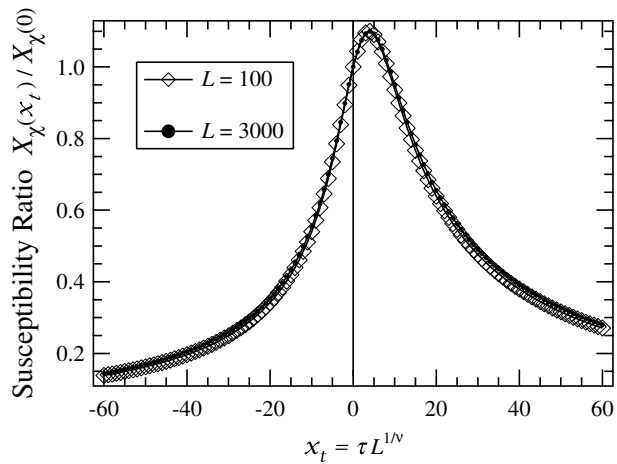}
\caption{The lines show the behavior of the scaling function of the
{\it normalized} susceptibility $X_\chi(x_t)/X_\chi(0)$ for a fluid
in which both the substrate-fluid and the fluid-fluid interactions
are {\it short-range}. The curves are for $L=100$ and  for $L=3000$.
The scaling variable is $x_t=\left(1-T_c/T \right)L^{1/\nu}$. Both
curves coincide perfectly well with each other. The main reason for
the deviation of the curves in Fig. \ref{100_3000_xi_sr_comp_best}
is the improper choice of the length $L$ (for a discussion see
Appendix \ref{scaling}).
  \label{100_3000_xi_sr_comp_norm_best}}
\end{figure}

The results from the finite-size scaling analysis of the behavior of
the susceptibility for a system with short-range interactions are
summarized in Figs. \ref{100_3000_xi_sr_comp_best} and
\ref{100_3000_xi_sr_comp_norm_best}. Fig.
\ref{100_3000_xi_sr_comp_best} compares the finite-size
susceptibilities $L^{-\gamma/\nu} \chi(T,h|L,h_{w,s})$ for
short-range films with $L=100$ and $L=3000$ layers. Fig.
\ref{100_3000_xi_sr_comp_norm_best} presents the corresponding
results for the ratio
$\chi(T,h|L,h_{w,s}=0)/\chi(T=T_c,h=0|L,h_{w,s}=0)$. In both
plots the scaling variable is $x_t=\left(1-T_c/T \right)L^{1/\nu}$.
The curves demonstrate a reasonably good scaling where the small
deviations for $L=100$ from the $L=3000$ curve can be explained with
the ambiguousness in the definition of $L$ (for a discussion see
Appendix \ref{scaling}), as well as with the corrections to scaling
terms and the role of the background (nonuniversal terms). Note that
the curves present the behavior of the {\it total} susceptibility
and not only of its singular part.

\subsection{The model with van der Waals type interactions}
\label{slri}

We note that the critical temperature $T_c$  depends on the
presence or absence of long-range fluid-fluid interactions in
the system. Let us denote by $T_{\rm c, lr}$ the critical
temperature of the system with subleading long-range fluid-fluid interactions (from van der
Waals type) and with $T_{\rm c, sr}$ the
corresponding temperature for short-range fluid-fluid interactions.
If $K=\beta J^l$ then it is straightforward to show that $K_{\rm c, lr}\simeq 0.161$, while $K_{\rm c, sr} \simeq 0.168$ in the
framework of our model defined by Eq.
(\ref{freeenergyfunctionalstarting}).

The results from the finite-size scaling analysis of the behavior of
the susceptibility for a system with long-range van der Waals type
interactions  are presented in Figs. \ref{100_3000_xi_lr_comp_best},
\ref{100_3000_xi_lr_comp_norm_best}, \ref{3000_lr_diff_h_comp},
\ref{3000_lr_diff_h_comp_xi}, \ref{3000_field_diff_t_comp} and
\ref{3000_field_diff_t_comp_xi}. The scaling procedure is explained
in details in  Appendix \ref{scaling}.

\subsubsection{The temperature dependence at $H=0$}

\begin{figure}[htb]
\includegraphics[width=\columnwidth]{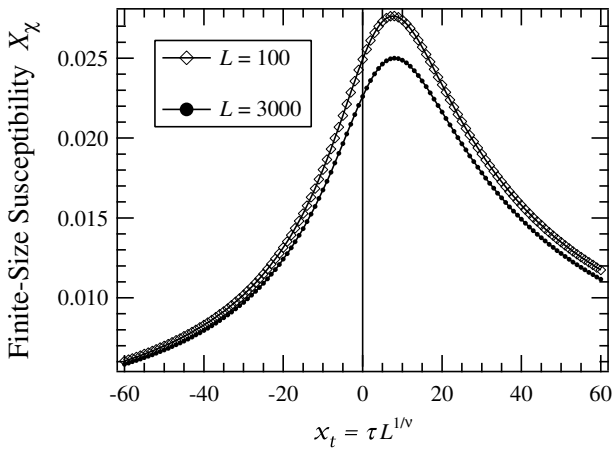}
\caption{The upper dot line (for $L=100$) and the lower dot line for
($L=3000$) show the behavior of the scaling function of the
susceptibility $X_\chi(x_t)$ for a fluid in which both the
substrate-fluid and the fluid-fluid interactions are of van der
Waals type, i.e. are {\it long-range}. The scaling variable is
$x_t=\left(1-T_c/T \right)L^{1/\nu}$. We observe that both
curves differ at most around their maximal value. The amplitude of
the surface field for $L=100$ is $h_{w,s}=0.73$ while for $L=3000$
it is $h_{w,s}=4$ which ensures that the variable
$2^{\sigma_s+1}|h_{w,s}|[L/\xi_0^+]^{\Delta/\nu-\sigma_s}=2^{4.5}
|h_{w,s}|[L/\xi_0^+]^{-1/2}$ has the same value for both the cases.
\label{100_3000_xi_lr_comp_best} }
\end{figure}

\begin{figure}[htb]
\includegraphics[width=\columnwidth]{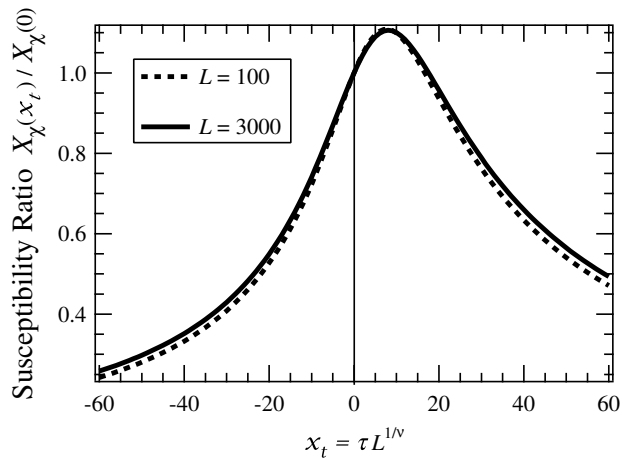}
\caption{The lines show the behavior of the scaling function of the
{\it normalized} susceptibility $X_\chi(x_t)/X_\chi(0)$ for a fluid
in which both the substrate-fluid and the fluid-fluid interactions
are {\it long-range}. The curves are for $L=100$ and  for $L=3000$.
 The scaling variable is $x_t=\left(1-T_c/T \right)L^{1/\nu}$.
  We observe that the both curves coincide perfectly well which each
  other only near the bulk critical point. The deviation of the curves
  from each other is due to the effect of the van der Waals
  interaction (compare with the short-range case) and increases with the increase
  of $|x_t|$. Note also that the scaling function decays much slower
  with $|x_t|$ in comparison with the short-range case.
  \label{100_3000_xi_lr_comp_norm_best}}
\end{figure}

Let us first consider the behavior of the finite-size susceptibility
in the absence of an external magnetic field, which is equivalent to
the behavior in a fluid system along the liquid-vapor coexistence
line. Our results are presented in Figs.
\ref{100_3000_xi_lr_comp_best} and
\ref{100_3000_xi_lr_comp_norm_best}. One observes that the
susceptibility possesses a maximum above the bulk critical
temperature (which reflects that fact that the $(+,+)$ boundary
condition stabilize the long-range order at temperatures slightly
above $T_c$) and that this maximum is weaker than for the
corresponding short-range system (see Fig.
(\ref{100_3000_xi_sr_comp_best})). The scaling variable is
$x_t=\left(1-T_c/T \right)L^{1/\nu}$. Note that the scaling
functions decay much more slowly as a function of $|x_t|$ in
comparison with the short-range case.  The maximum of the
short-range case is around $x_t \simeq 4$ while in the case of a van
der Waals type system it is around $x_t \simeq 8$. These results
imply that, as expected, the long-range tails of the interactions
help to stabilize the long-range order even a bit above the
corresponding limit in $T$ for the short-range system with $(+,+)$
boundary conditions.

\subsubsection{The temperature dependence of the susceptibility at
$H\ne 0$}

\begin{figure}[htb]
\includegraphics[width=\columnwidth]{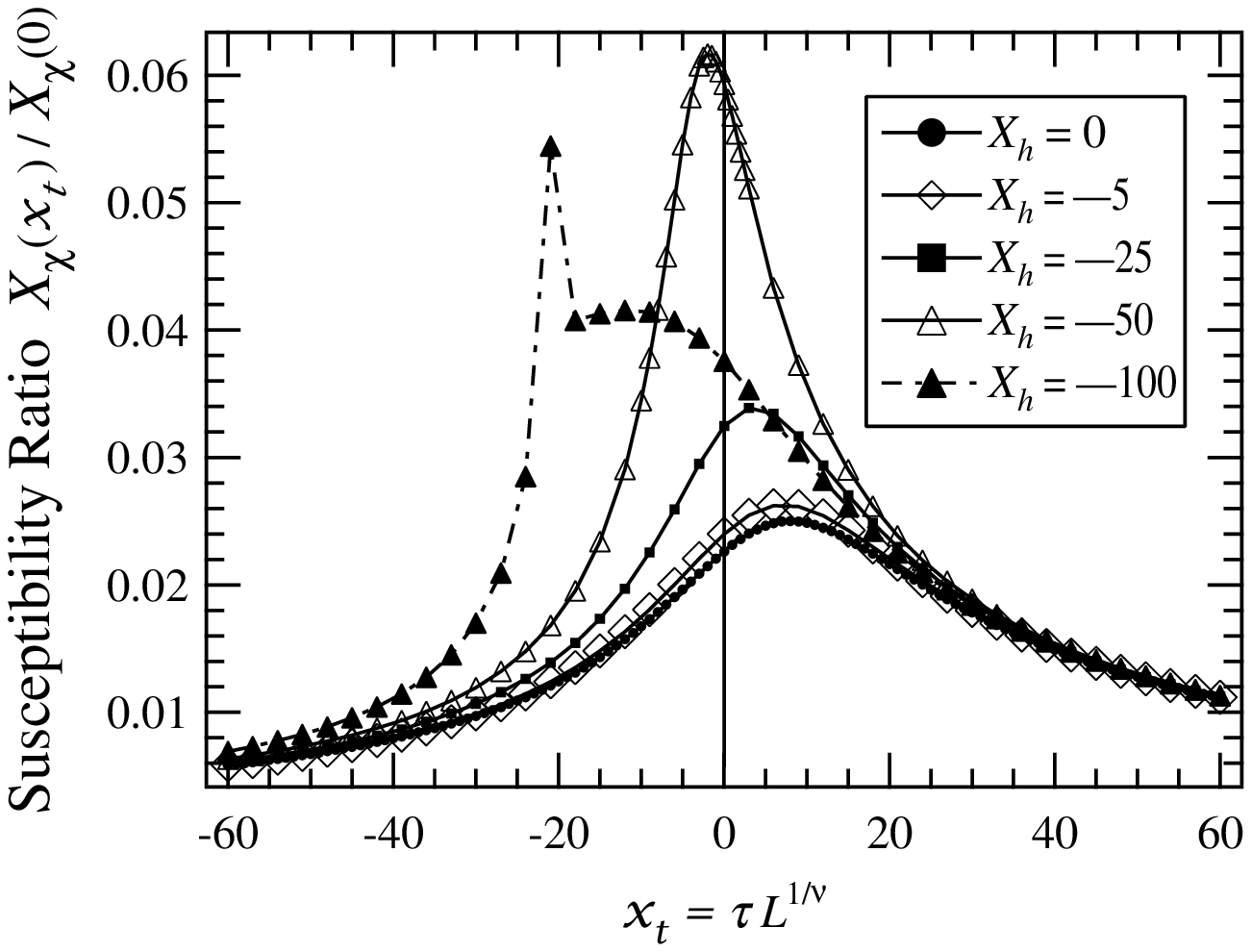}
\caption{The lines show the behavior of the scaling function of the
susceptibility $X_\chi(x_t,x_h)$ for a fluid in which both the
substrate-fluid and the fluid-fluid interactions are {\it
long-range}. The substrate-fluid potential is characterized by
$h_{w,s}=4$, which correspond to the situation of $^3$He or $^4$He
films bounded by Au surfaces (see Appendix \ref{helium}). The curves
are for $L=3000$ and at $x_h=0, -5, -25, -50, -100$, where
$x_h=\beta H L^{\Delta/\nu}$. The scaling variable is
$x_t=\left(1-T_c/T \right)L^{1/\nu}$. The changed shape of the curve
for $x_h=-100$ signals that this curves is the precursor of a
first-order phase transition.
  \label{3000_lr_diff_h_comp}}
\end{figure}

\begin{figure}[htb]
\includegraphics[width=\columnwidth]{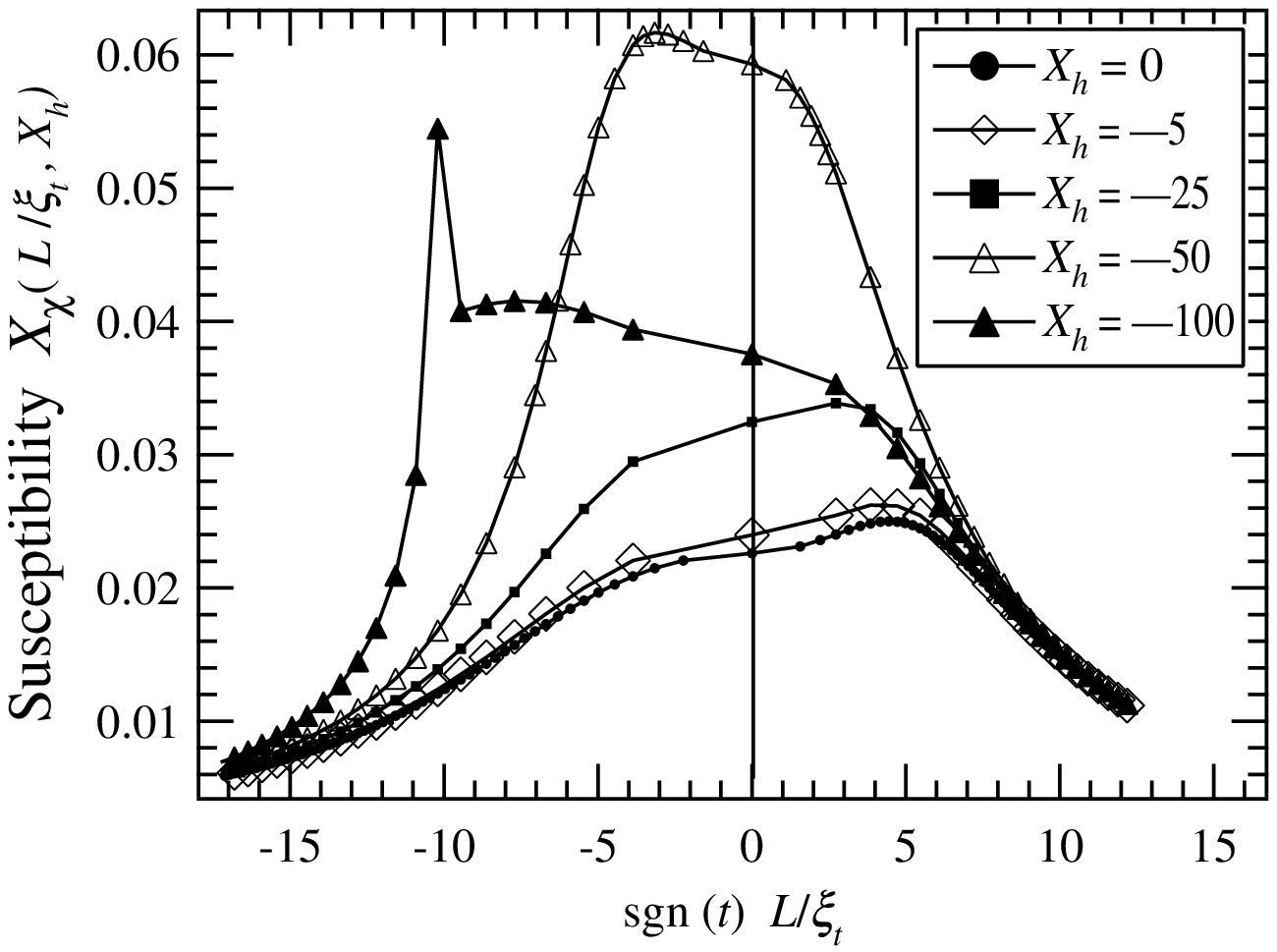}
\caption{The same as in Fig. \ref{3000_lr_diff_h_comp} but the
scaling function is presented as a function of $L/\xi_t$ at several
fixed values of $x_h$.  We expect this curve to be a good
approximation of the corresponding finite-size scaling function for
a real three-dimensional experimental system of $^3$He or $^4$He
films bounded by Au surfaces.
  \label{3000_lr_diff_h_comp_xi}}
\end{figure}

In this section we consider the behavior of the finite-size
susceptibility for values of the external bulk magnetic field that
support the vapour phase of the fluid. The results are presented in
Fig. \ref{3000_lr_diff_h_comp}. Note that by changing the sign of $H$ (by choosing negative values of $x_h=\beta H L^{\Delta/\nu}$) one can
show that the position of the maximum in the behavior of the
susceptibility, which for $H=0$ is at $T>T_c$, moves gradually
toward $T_c$ and for negative field with large enough magnitude can
even be {\it below} $T_c$.
In Fig. \ref{3000_lr_diff_h_comp_xi} we present the above results
as a function of $L/\xi_t$. This resolves the question about the
value of the nonuniversal metric factor in $x_t$. Note that then the
shape of the curve and the position of its maximum shall be a
reasonable approximation for the real experimental system of $^3$He
or $^4$He films bounded by Au surfaces.

\subsubsection{The field dependence of the susceptibility at $T\ne T_c$}

\begin{figure}[htb]
\includegraphics[width=\columnwidth]{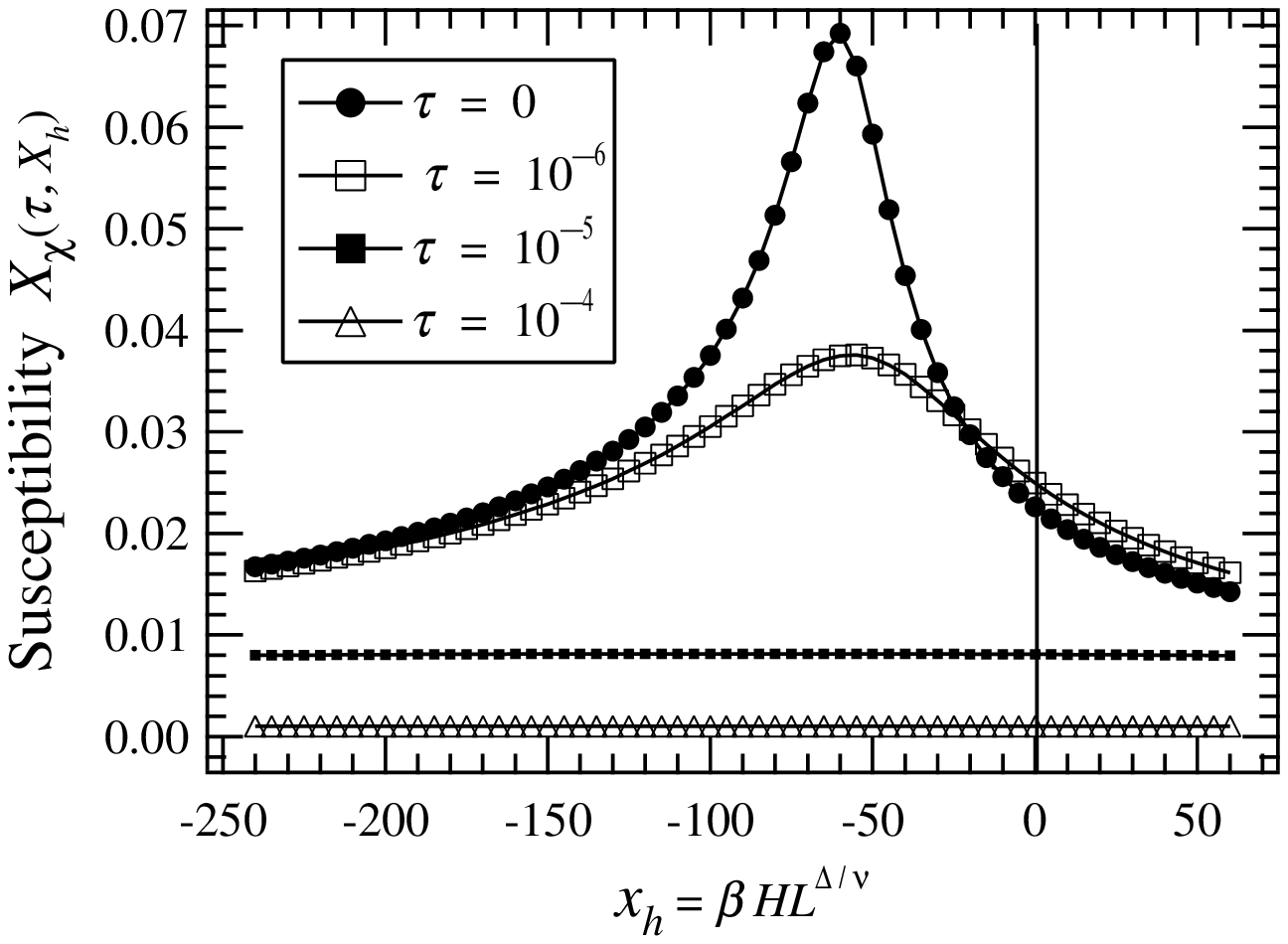}
\caption{The lines show the behavior of the scaling function of the
susceptibility $X_\chi(x_t,x_h)$ as a function of $x_h$ at several
fixed values of $x_t$ for a fluid in which both the substrate-fluid
and the fluid-fluid interactions are {\it long-range}. The
substrate-fluid potential is characterized by $h_{w,s}=4$, which
correspond to the situation of $^3$He or $^4$He films bounded by Au
surfaces (see Appendix \ref{helium}). The scaling variable is
$x_h=\beta H L^{\Delta/\nu}$. The numerical calculations are
performed for $L=3000$ layers.
  \label{3000_field_diff_t_comp}}
\end{figure}

\begin{figure}[htb]
\includegraphics[width=\columnwidth]{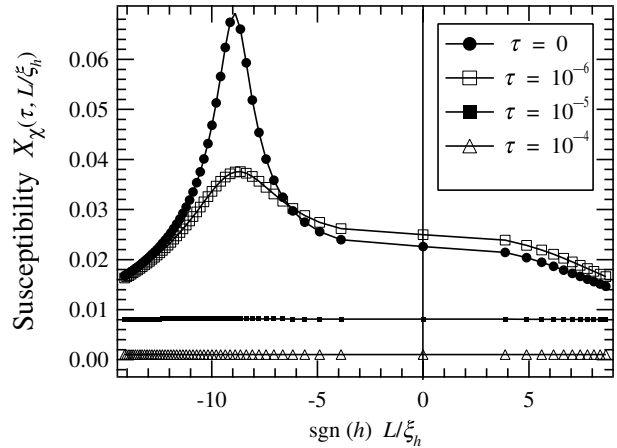}
\caption{The same as in Fig. \ref{3000_field_diff_t_comp} but the
scaling function is presented as a function of $L/\xi_h$ at several
fixed values of $\tau$.  We expect this curve to be a good
approximation of the corresponding finite-size scaling function for
a real three-dimensional experimental system of $^3$He or $^4$He
films bounded by Au surfaces.
  \label{3000_field_diff_t_comp_xi}}
\end{figure}

Here we analyze the behavior of the finite-size susceptibility as a
function of the field scaling variable $x_h=\beta H L^{\Delta/\nu}$
for fixed values of the temperature close to the bulk critical
temperature $T_c$. The results are presented in Fig.
\ref{3000_field_diff_t_comp}. The curves are for $
\tau=0,10^{-6},10^{-5},10^{-4}$. Note that
the behavior of the susceptibility possesses a peak that essentially
differs from the corresponding background contribution only for
$|\tau| \lesssim 10^{-6}$.
In Fig. \ref{3000_field_diff_t_comp_xi} we present the same results,
this time as a function of $L/\xi_h$. This resolves the question
about the value of the possible nonuniversal metric factor in $x_h$.
Thus, the shape of the curve and the position of its maximum  shall
be a reasonable approximation for the real experimental system of
$^3$He or $^4$He films bounded by Au surfaces.

\section{Concluding remarks and discussion}

In this article we investigated the behavior of the susceptibility
in thin films with van der Waals type long-range interactions.
Prominent examples of such systems are simple nonpolar fluids in
thermodynamic equilibrium with their vapor, as well as binary fluid
mixtures close to their demixing point. We have seen that despite
the fact that this kind of interaction does not change the critical
exponents of the system it nevertheless gives rise to a variety of
finite-size effects that become dominant when $L/\xi_\infty \gg 1$.
Furthermore, we have formulated a criterion regarding the conditions
under which such effects also essentially modify the finite-size
behavior of the susceptibility everywhere within  the critical
region. According to this criterion  if the thickness $L$ of the
film is such that (see Eq. (\ref{cL}))
\begin{equation}\label{cdiscusiion}
L \lesssim L_{\rm crit}=\xi_0^+(16 |h_{w,s}|)^{\nu/\beta},
\end{equation}
the effects due to the van der Waals interaction are essential
and cannot be neglected. Here $\xi_0^+$ is the amplitude of the
bulk correlation length while the (dimensionless) factor $h_{w,s}$
characterizes the strength of the fluid-substrate interaction $\sim
h_{w,s} z^{-\sigma}$, where $z$ is the distance from the substrate
surface toward to the bulk of the fluid. In the case of $^3$He or
$^4$He bounded by Au surfaces we find that at the
corresponding critical point of both the fluids $h_{w,s}\simeq 4$
(see Appendix \ref{helium}). Then, for a three-dimensional Ising
type system the direct evaluation yields the estimate $L_{\rm
crit}\simeq 9000$\;{\AA}. A comparison when $L/a=3000$ layers
(where $a$ is the lattice constant) between the behavior of the
finite-size susceptibility of a system with completely short-range
interactions and one with long-range fluid-fluid and
substrate-fluid interactions is given in Fig.
\ref{3000_sr_lr_comp.eps}. One observes a clear distinction between
the curves in the critical region. The calculations were preformed using
a mean-field type model which is described in detail in
Section \ref{model}. In order to determine the susceptibility one
solves $L/a=3000$ coupled nonlinear equations; we make use of the
Newton-Kantorovich method. We have chosen a film with this number
of layers because it corresponds to a realistic experimental setup
of $^3$He film between the Au electrodes of an experimental cell
in which the smallest distance between the plates is $1.5\;\mu$m;
the distance $r_0$ between the atoms of $^3$He at its
liquid-vapor critical point has been estimated to be $r_0\simeq 4.9
\; {\AA}$ (see Appendix \ref{helium}). For such a system the
behavior of the susceptibility as a function of $(1-T_c/T)L^{1/\nu}$
for different values of the bulk field $x_h=\beta H L^{\Delta/\nu}$
is shown in Fig. \ref{3000_lr_diff_h_comp}, while in Fig.
\ref{3000_field_diff_t_comp} the susceptibility is plotted as a
function of $x_h$ for a few fixed values of $t$. The same data are
shown in Figs. \ref{3000_lr_diff_h_comp_xi} and
\ref{3000_field_diff_t_comp_xi}
 as functions of $L/\xi_t$ and
$L/\xi_h$, respectively. We expect these curves to resemble the
actual experimental data for $^3$He (or $^4$He) film. One observes
that for $x_h=0$, the
maximum of the susceptibility is {\it above} the bulk critical
point. Furthermore, we note that when one is on the vapor side of
the bulk phase diagram the maximum moves toward the bulk critical
point with increasing distance from the coexistence curve in the
phase diagram as a function of $-x_h$  and that for $-x_h$ large
enough the maximum positions itself at a temperature less than
$T_c$. Additionally, one observes that as a function of the scaling
field variable ($x_h$, or $y_h=L/\xi_h$) the susceptibility does not
display an easily distinguishable maximum unless $|t| \lesssim
10^{-6}$. The position of this maximum changes very slightly with
$t$ and is around ${\rm sign}(h) L/\xi_h \simeq -8.7$ on the vapor
side of the bulk phase diagram.

We stress that all of the above curves depend on the value of
$h_{w,s}$ and are, thus, nouniniversal. This implies that if
$h_{w,s}$ is kept fixed when $L$ changes one will obtain different
curves for the different $L$'s. The same will also be true when $L$
is kept fixed but $h_{w,s}$ changes. This is illustrated in Appendix
\ref{scaling}---see Figs. \ref{role_of_hs} and
\ref{hs_fixed_diff_L}. Only when $h_{w,s}/\sqrt{L}$ is kept fixed can a curve
that does not depend on $L$ nor on $h_{w,s}$, but just on
their ratio, be obtained - see Fig. \ref{role_of_L}. In practice, one does not know the precise
value of the system size $L$ - an important issue for
very thin film thicknesses (for more details see Appendix
\ref{scaling}). When this is also taken into account further
improvement of the data collapse can be achieved as seen in Fig.
\ref{role_of_L_norm}, where the corresponding data for the
susceptibility for fixed $h_{w,s}/\sqrt{L}$ are  normalized by its
value at the bulk critical point. Definitely the above predictions
are clearly experimentally verifiable. For a given fluid, one can either change the size $L$ of the distance between the
plates of the experimental cell (i.e. the fluid film) or
the corresponding substrate (i.e. the value of $h_{w,s}$). The
bigger the change in the new $h_{w,s}$, the bigger will
be the deviation of the new curve of the finite-size susceptibility
from the old one. Of course, this will be possible only if
$L\lesssim L_{\rm crit}$. {\it When $L\gg L_{\rm crit}$ the
universal behavior of the finite-size susceptibility} in its
standard form {\it will prevail} and the van der Waals interaction
will only lead to small, probably experimentally negligible,
corrections to the {\it universal} curve. Let us also note that
when the experiment is performed for $^3$He or $^4$He around their
liquid-vapor critical points according to our predictions one will
obtain for any fixed $L$ -- even when $L\lesssim L_{\rm crit}$ --
practically the same curve $X_\chi$ for the finite-size
susceptibility of both the fluids. For $L\lesssim L_{\rm crit}$ this
will be simply due to the fact that $h_{w,s}\simeq 4$ for {\it both}
the fluids (see Appendix \ref{helium}), while for $L\gg L_{\rm
crit}$ that will be due to universality since near their liquid
vapor critical points these fluids belong to the
three-dimensional Ising universality class.

In the work described above, the effects of retardation on the van
der Waals force have been neglected. These effects set in for
distances $r$ larger that $160$ {\AA} between the He and Au atoms
\cite{CC88}. A possible concern is the influence of retardation on
the system   investigated in this paper.
We have estimated that influence by performing numerical
calculations in which an ``extreme'' retardation was imposed, in
that the interaction potential was set to zero for separations
$r>100$ layers. We discovered that the numerical consequence of
retardation grows with increasing  film thickness. At  $L=3000$ the
difference between the calculation with the unretarded van der Waals
force and the force with long-distance cutoff is  $13.5 \%$ at
$T=T_c$. This is an overestimate of the actual influence of
retardation in experimental realizations. We note that retardation
reduces the suppression of the susceptibility by the van der Waals
interaction---see Fig. \ref{3000_sr_lr_comp.eps}.

Finally, let us note that in the current study we did not take into
account the influence of gravity on the behavior of the
finite-size susceptibility. Because gravity leads to
stratification of the density one expects an additional, more
complicated $z$ dependence of the local and, therefore, of the total
susceptibility. We hope to return to this question in a future
publication.

\acknowledgments

A portion of this work was carried out under a contract with the
National Aeronautics and Space Administration.

D.D. acknowledges the partial financial support under grant F-1402
of the Bulgarian NSF.

\appendix
\section{Derivation of the  equation for the
magnetization profile in a van der Waals film}
\label{profiles}

Here we  derive an explicit form for the function $\hat{{\cal
J}}(z)$ for some basic cases of special physical interest. Namely we
take $J({\bf r})$ to be of the "van der Walls form"
\begin{equation}\label{defJ}
    J({\bf r})=\frac{J}{1+r^{d+\sigma}},
\end{equation}
where $r=|{\bf r}|$, and $d=\sigma=3$ for the "real" (nonretarded)
van der Waals interaction.

 Then one can further simplify the sum in the right-hand
side of the above equation. Using the identity
\begin{equation}\label{id}
\frac{1}{1+z^\alpha}=\int_0^\infty dt \ \exp(-z t) t^{\alpha-1}\
E_\alpha,_\alpha (-t^\alpha),
\end{equation}
where
\begin{equation}
\label{ML} E_\alpha,_\beta (z)=\sum_{k=0}^\infty
\frac{z^k}{\Gamma(\alpha k +\beta)}, \qquad \alpha>0,
\end{equation}
 are the Mittag-Leffler functions, the sum can be
rewritten in the form
\begin{eqnarray}\label{sum1}
\sum_{{\bf r}'}J({\bf r}-{\bf r}')m(z')&=& \sum_{z'=0}^{L}
\hat{{\cal J}}_{d,\sigma}(z-z')m(z'),
\end{eqnarray}
where
\begin{eqnarray}\label{Gds}
\lefteqn{\hat{{\cal J}}_{d,\sigma}(z)} \nonumber \\ &=&J \int_0^\infty dt \
t^{(d+\sigma)/2-1}\
E_{\frac{d+\sigma}{2},\frac{d+\sigma}{2}}(-t^{(d+\sigma)/2}) \nonumber \\ && \left(\sum_{{r}_\|}e^{-t{\bf
r}_\|^2}\right) e^{-tz^2}.
\end{eqnarray}
The main advantage of the above form is that it factorizes the
summation over the components of ${\bf r}_\|'$. With the help of the
Poisson identity Eq. (\ref{sum1}) becomes
\begin{widetext}
\begin{eqnarray}
\label{sumgenfinal} \sum_{{\bf r}'}J({\bf r}-{\bf r}')m(z')&=&
J\left(\sum_{{\bf r}_\|'}\frac{1}{1+{\bf
r}_\|'^{d+\sigma}}\right)m(z)\\
& & + J\int_0^\infty dt \ \left(\frac{\pi}{t}\right)^{(d-1)/2}
t^{\frac{d+\sigma}{2}-1}\
E_{\frac{d+\sigma}{2},\frac{d+\sigma}{2}}\left(-t^{\frac{d+\sigma}{2}}\right)
\sum_{z'=0 \atop z\ne z'}^{L}e^{-t(z-z')^2} m(z')\nonumber \\
& & + J\int_0^\infty dt \ \left(\frac{\pi}{t}\right)^{(d-1)/2}
t^{\frac{d+\sigma}{2}-1}\
E_{\frac{d+\sigma}{2},\frac{d+\sigma}{2}}\left(-t^{\frac{d+\sigma}{2}}\right)\sum_{{\bf
n} \in {\mathbb Z}^{d-1} \atop {\bf n}\ne {\bf 0}}e^{-\pi^2{\bf
n}^2/t}\sum_{z'=0\atop z\ne z'}^{L}e^{-t(z-z')^2} m(z'), \nonumber
\end{eqnarray}
\end{widetext}
where ${\bf n}\in {\mathbb Z}^{d-1}$  is a $(d-1)$-dimensional
vector with integer components, and all the lengths are measured in
units of lattice spacings. It is easy to show that
$\max_{t}\exp(-\pi^2 {\bf n}^2/t+t(z-z')^2)$ is attained at $t=\pi
|{\bf n}|/|z-z'|$ and is equal to $\exp({-2\pi |{\bf n}| |z-z'|})$.
Because of this exponential decay in the last row of the above
equation we will take into account only the terms with $|{\bf n}|=1$
and $|z-z'|=1$ (the corresponding improvements that take into
account ${\bf n}=2, 3, \cdots$ and $|z-z'|=2, 3, \cdots$ are
obvious; as we will see even the contributions stemming from $|{\bf
n}|=1$ and $|z-z'|=1$ are very small). It is clear that size
dependent contributions that are due to the terms in the last row of
(\ref{sumgenfinal}) will be exponentially small in $L$.

For $d=\sigma=3$ the corresponding Mittag-Leffler function can be
expressed in the following simple form
\begin{eqnarray}\label{ML3}
E_{3,3}(-t^3)&=& \frac{1}{3
t^2}\left[e^{-t}-2e^{t/2}\cos\left(\frac{\pi}{3}+\frac{\sqrt{3}}{2}t
 \right)\right].\;\;\;\;
\end{eqnarray}

Taking into account that, if $x>0$,
\begin{equation}
\pi \int_0^\infty t E_{3,3}(-t^3)e^{-t x}dt ={\cal G}_3(x),
\end{equation}
where
\begin{eqnarray}
\label{g} {\cal G}_3(x)&=&\frac{\pi}{3}\left[\sqrt{3}\ \arctan{\Big|
\frac{\sqrt{3}}{2x-1}}\Big|-\ln\left(1+\frac{1}{x}\right)\right.\nonumber
\\
&&
\left.+\frac{1}{2}\ln\left(1-\frac{1}{x}+\frac{1}{x^2}\right)\right],
\end{eqnarray}
we arrive at the following equation for the magnetization profile
\begin{widetext}
\begin{equation}
\label{d3em} {\rm arctanh}\left[ m^*(z)\right]= h(z)+K\left\{c_2
m^*(z) + c_2^{nn} \left[m^*(z+1)+m^*(z-1)\right]+\sum_{z'=0 \atop
|z'- z|\ge 2}^{L}{\cal G}_3(|z-z'|^2)m^*(z')\right\},
\end{equation}
\end{widetext}
where
\begin{equation}\label{c2}
c_2=\sum_{{\bf n}\in {\mathbb Z}^2}\frac{1}{1+|{\bf n}|^6}\simeq
3.602,
\end{equation}
and
\begin{eqnarray}\label{c2nn}
c_2^{nn}&=& - \frac{8}{3}\pi \int_0^\infty e^{-t/2-
\pi^2/t}\cos\left(\frac{\sqrt{3}}{2}t+\frac{\pi}{3}\right) +{\cal G}_3(1) \nonumber \\
&=&-\frac{8}{3} \pi \left[(-1)^{1/3}K_0(\sqrt{2-2i\sqrt{3}\pi}) \right. \nonumber \\
&& \left.-(-1)^{2/3}K_0(\sqrt{2+2i\sqrt{3}\pi})\right]
+ \frac{\pi}{3}\left(\frac{\pi}{\sqrt{3}}-\ln 2\right)\nonumber\\
&\simeq & 0.00955+ 1.17354 \approx 1.183.
\end{eqnarray}
As we see, the contribution due to the first part in
Eq.\,(\ref{c2nn}), and therefore the contributions due to the last
row of Eq.\,(\ref{sumgenfinal}) are of the order of $1\%$ in the
constant $c_2^{nn}$. It is easy to verify that
\begin{equation}
\label{assg} {\cal G}(x)\simeq \frac{\pi}{2}x^{-2}-\frac{\pi}{5}
x^{-5}+\frac{\pi}{8} x^{-8}+O(x^{-11}), \qquad x\rightarrow
\infty.
\end{equation}
Setting  ${\cal G}\equiv 0$ in Eq. (\ref{d3em})  one obtains an
equation having a form that is familiar in the mean-field theory
of short-range systems. Actually in our case the system in question
has short-range interactions in $z$ direction and long-range
ones within the planes perpendicular to $z$. The standard
Ginzburg-Landau equation follows, for small $m$, after taking into
account that ${\rm arctanh}(m)\simeq m + m^3/3 +O(m^5)$. A
continuum version of the equation follows from the replacement
$m(z-1)+m(z+1)\rightarrow 2m(z)+m''[z]$. Obviously such a
continuum version can also be constructed for the long-range
system by adding the terms contributed by the function
${\cal G}(x)$, which is, in this case, a continuous
function.
\begin{figure}
\includegraphics[width=\columnwidth]{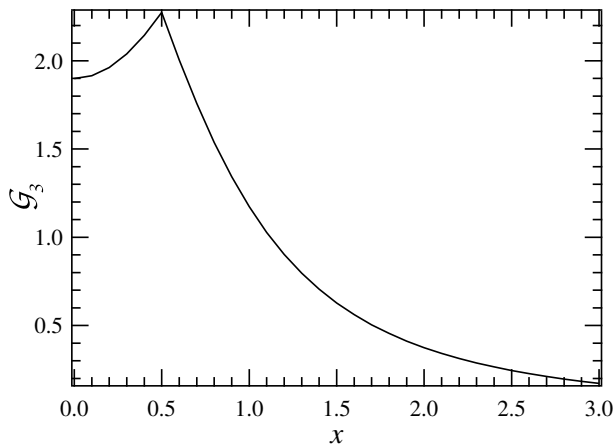}
\caption{The behavior of the ${{\cal G}_3}$-function as defined in (\protect{\ref{g}}). \label{gplot}}
\end{figure}
Note that the function ${\cal G}(x)$ is well defined everywhere
for $x\ge 0$ and not only for $x\ge 1$ as we actually need it in
the lattice formulation of the theory. Thus, in the continuum
formulation of the theory the integration can be extended over the
region $z\in [0,L]$. This does not change the long-range behavior
of the magnetization profiles. In the continuum case
the equation for the magnetization profile reads
\begin{widetext}
\begin{equation}\label{eqcont}
m^*[z]+\frac{1}{3}(m^{*}[z])^3=h[z]+K\left\{c_2 m^*(z) + c_2^{nn}
\left[2 m^*[z]+\frac{d^2 m^*[z]}{dz^2}\right]+\int_{0}^{L}{\cal
G}(|z-z'|^2)m^*(z')dz' \right\}.
\end{equation}
\end{widetext}

\section{How to scale}
\label{scaling}

\begin{figure}[htbp]
\includegraphics[width=\columnwidth]{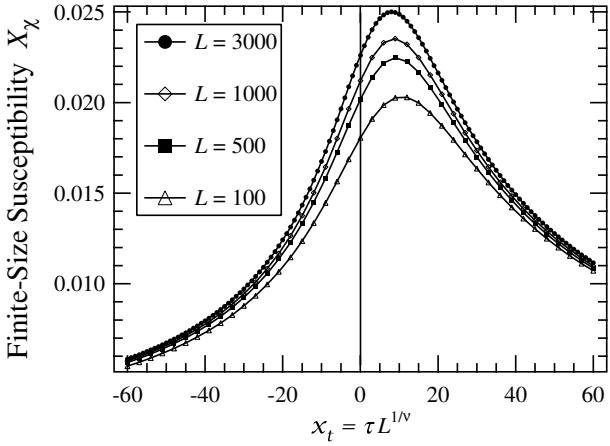}
\caption{The behavior of the finite-size  susceptibility normalized
per $L^{\Delta/\nu}$ for different fixed values of $L$ with the
surface-field amplitude kept fixed at $h_{w,s}=4$.  Systems with
thickness  $L=3000,1000,500$ and $L=100$ are considered.  The curves
represent the behavior of the susceptibility of $^3$He and $^4$He
films with different thickness when the surrounding surfaces are
made of gold. \label{hs_fixed_diff_L}}
\end{figure}

\begin{figure}[htbp]
\includegraphics[width=\columnwidth]{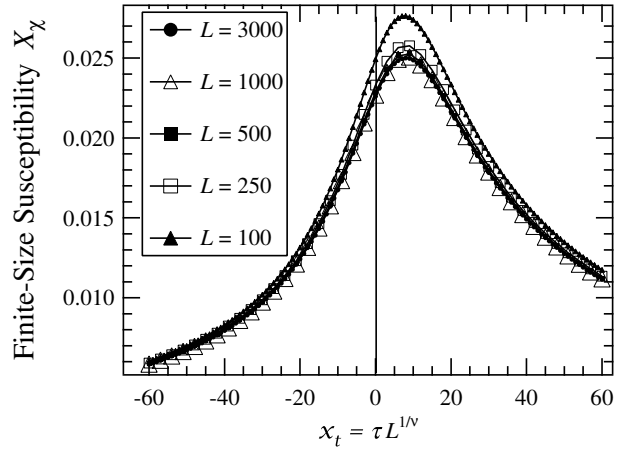}
\caption{The finite-size susceptibility for $L=3000$, $L=1000$,
$L=500$, $L=250$, $L=100$. The amplitude of the surface magnetic
field is rescaled in such a way that $h_{w,s}/\sqrt{L}={\rm const}$
for all values of $L$ and, as in the experimental realization,
$h_{w,s}=4$ for $L=3000$. One observes that practically all the
curves for $L\ge 250$ coincide with each other, i.e. the scaling is
indeed valid. The curve with $L=100$ differs from the others. Thus
$L=100$ is too small, and in this case the importance of
$\delta$ is demonstrated.  \label{role_of_L}}
\end{figure}

\begin{figure}[hb]
\includegraphics[width=\columnwidth]{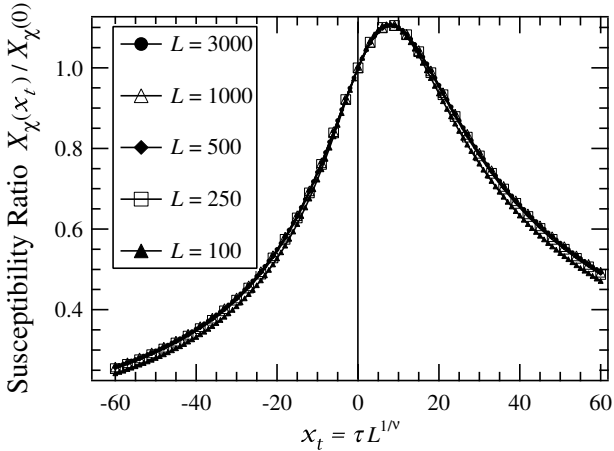}
\caption{The finite-size susceptibility for $L=3000$, $L=1000$,
$L=500$, $L=250$, and $L=100$ normalized to its value at
$T=T_c$. The amplitude of the surface magnetic field is rescaled in
such a way that $h_{w,s}/\sqrt{L}={\rm const}$ for all values of
$L$. \label{role_of_L_norm}}
\end{figure}

In the discussion up to now we have tacitly assumed that the value
of $L$ is precisely known. This is, however, not only an
experimental problem---due to the roughness  of the surface, the
existence of impurities, dust, etc.---but also a theoretical problem
that might play a role when $L$ is not ``large enough". Let us make a
brief comment about this issue.  The definition of the size of the
system is unambiguous only for systems with periodic boundary
conditions. If $N$ is the number of layers with independent degrees
of freedom, then the size of the system is simply $L=Na$, where $a$
is the distance between the layers. Any point in the system is
equivalent to any other point. Therefore, any layer is equally
suitable to be taken as the origin with respect to which one measures distance.
However, how one proceeds for a system with $(+,+)$ boundary
conditions---when the first and the last layers have fixed degrees
of freedom---is less clear. For consistency with periodic boundary conditions one
can, of course, count the number of layers with independent
degrees of freedom and let this, as in the case of periodic boundary conditions, be equal to $N$. Then the question
is, shall we include in the total size of the system the half
distance between the two outermost layers with independent degrees
of freedom and the adjacent layers with the fixed degrees of
freedom? A reasonable approach seems to be that at least half of
these distances should be taken to belong to the system, i.e.
$L=(N+1)a$. This, of course, is not an unambiguous procedure. We
have two layers with fixed degrees of freedom, which do not belong
to the ``substrate" surrounding the system that strongly prefers the
ordered phase of the system in the case of $(+,+)$ boundary
conditions. Thus one has to somehow decide which portion of the
bounding layers are to be counted within the system. Much more
complicated is the case of systems with long-range interactions.
Then any particle (atom, or molecule) of the system interacts with
any other one from the substrate. How then does one define the size
of the system or the borderline between the substrate and the
system? Let as denote by $\hat{L}$ the ``true" size of the system
(which we do not know), and by $L$ a reasonable approximation of
that size (say, by taking, as above for the short-range case, the
size to be $L=(N+1)a$). The last implies that $\hat{L}=L+\delta$,
where $\delta \ll \hat{L}$.
\begin{figure}[htbp]
\includegraphics[width=\columnwidth]{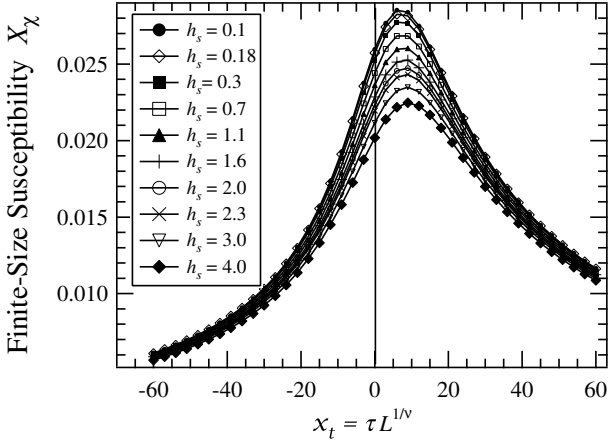}
\caption{The behavior of the finite-size  susceptibility normalized
per $L^{\Delta/\nu}$ for different fixed values of $h_{w,s}$ and for
$L=500$. Within the mean-field approximation one has that
$\Delta/\nu=3$. The values of $h_{w,s}$ for which results are
presented are $h_{w,s}=0.1,0.18,0.3,0.7,1.1,1.6,2.0,2.3,3.0$ and
$4.0$. They model the role of different substrates that surround a
given fluid film with thickness $L$. \label{role_of_hs}}
\end{figure}
Figure \ref{role_of_hs} demonstrates the difficulties when studying
the scaling in systems with subleading long-range interactions of
the van der Waals type. The role of the long-range surface
potentials, which are {\it irrelevant} in the renormalization group
sense but for moderate values of $L$, i.e. for $a \ll L \ll L_{\rm
crit}$ (see Eq. (\ref{cL})), contribute to the {\it leading}
behavior of the finite-size susceptibility is clearly seen. One can
say that, for such values of $L$, the quantity $x_s$ is a sort of ``dangerous"
irrelevant variable---in the sense that, despite being irrelevant,
one \emph{cannot} neglect it when $L<L_{\rm crit}$. We further
note that the greatest deviation of the curves for different $L$
from each other is around the maximum value of the scaling
functions. The lack of the data collapse is due to the fact that
$x_s \sim h_{w,s}/\sqrt{L}$ is {\it not} the same for all the curves
(see Eq. (\ref{cL})). Definitely similar spreading of the data for
the finite-size susceptibility are to be observed if $h_{w,s}$ is
kept fixed while $L$ changes---say $h_{w,s}=4$ as in case of $^3$He
or $^4$He confined by Au plates and---one considers
$L=3000,1000,500,100$ (see Fig. \ref{hs_fixed_diff_L}).
However, when $x_s$ is kept fixed for the same values of $L$
considered before, the data collapse improves greatly, as shown in
Fig. \ref{role_of_L}.  One observes that violations of scaling are
now clearly detectable only for the smallest system size $L=100$.
Let us now recall that when $L$ is Ê``large but small enough" the
effect of $\delta$, i.e. of the fact that we do not know exactly the
system size $L$ is clearly evident. In the case of $\chi$ one has
$\chi \simeq \hat{L}^{\gamma/\nu} [1+\delta (\gamma/\nu)
\hat{L}^{-1}+O(\delta^2 \hat{L}^{-2})]
X_\chi(\hat{x}_t,\hat{x}_h|\hat{x}_s,\hat{x}_b,\hat{x}_\omega)$,
where $\hat{x}_t=a_t t \hat{L}^{1/\nu}$, $\hat{x}_h=a_h h
\hat{L}^{\Delta/\nu}$, $x_s=h_{w,s} \hat{L}^{(d+2-\eta)/2-\sigma}$,
$\hat{x}_b=b\hat{L}^{2-\eta-\sigma}$, $\hat{x}_\omega=a_\omega
\hat{L}^{-\omega}$. Note now that the expansion of the above
expressions in terms of $\hat{L}$ will yield all possible
non-universal (proportional to $\hat{L}^{-1}$) corrections to the
leading finite-size behavior of the susceptibility with the greatest
deviation of the curves for different $\delta$ occurring near the
maximum value of $X_\chi$. For the sake of precision let us also
note that similar corrections will be produced if one takes into
account the change from $L$ to $\hat{L}$ in the variables
$\hat{x}_{t}, \hat{x}_h, \hat{x}_s, \hat{x}_b$ and $\hat{x}_\omega$.
Thus, only the leading finite-size behavior can be determined
unambiguously. All corrections will depend on the definition of $L$,
i.e. on $\delta$. There is, nevertheless, still something that one
can do in order to check that the behavior of the susceptibility for
$L=100$ is simply due to the above explained unambiguity in the
definition of $L$. Note, that if we normalize $\chi$ to its value at
a given point within the critical region -- say to the value
$\chi_0$ at the bulk critical point ($T=T_c$, $H=0$) -- then,
whatever the definition of $L$ is, the {\it leading} behavior of the
resulting quantity will {\it not} depend on this definition.
Explicitly, we have
\begin{equation}
\frac{\chi}{\chi_0}\simeq
\frac{X_\chi(\hat{x}_t,\hat{x}_h|\hat{x}_s,\hat{x}_b,\hat{x}_\omega)}
{X_\chi(0,0|\hat{x}_s,\hat{x}_b,\hat{x}_\omega)}+O(\hat{L}^{-1}).
\end{equation}
The result from the application of the above procedure is shown in
Fig. \ref{role_of_L_norm}.
We observe that all the curves, including $L=100$, now allow for data
collapse and that only a small deviation of the curves from each
other is observed for very large values of the scaling
variable $x_t$, when the onset of the nonuniversal corrections to
scaling due to the role of the fluid-fluid interaction (i.e.
proportional to $x_b$) is expected to set in. Thus, despite  ignorance
of the precise value of $L$, we are able to determine the
leading finite-size behavior of the susceptibility.

\section{Evaluation of the scaling field parameters for Helium}
\label{helium}

We imagine a simple model for  helium ($^3$He or $^4$He) in which
atoms interact via a pair potential $w^l({\bf r},
{\bf r}')=-4J^l({\bf r}, {\bf r}')$. We assume that the fluid is
bounded by a substrate whose particles interact with the helium
particles with a pair potential $w^{l,s}({\bf r}, {\bf
r}')=-4J^{l,s}({\bf r}, {\bf r}')$. Within the lattice gas model for
any given configuration ${\cal C}$ of particles $\{p_i^s,p_j^l\}$,
$i \in {\cal S}$, $j \in {\cal L}$ with ${\cal L}$ and ${\cal S}$
denoting the region occupied by the helium and substrate particles,
respectively, the energy of the fluid is given by
\begin{eqnarray}
{\cal E} &=& \sum_{i\in {\cal S},j\in {\cal
L}}w_{i,j}^{l,s}p_{i}^{s}p_{j}^{l} + \frac{1}{2}\sum_{i,j\in {\cal
L}}w_{i,j}^{l}p_{i}^{l}p_{j}^{l}\\ \nonumber
&=&-4\sum_{i\in {\cal
S},j\in {\cal L}}J_{i,j}^{l,s}p_{i}^{s}p_{j}^{l} -{2}\sum_{i,j\in
{\cal L}}J_{i,j}^{l}p_{i}^{l}p_{j}^{l},
\end{eqnarray}
where $p_j^l\in \{0,1\}$ and $p_i^s \in \{0,1\}$ denote the
occupation numbers for the fluid and substrate  particles,
respectively. Since only the part $\{p_j^l\}$ belonging to the fluid
becomes critical around $T_c$ and the fluctuations of the particles
$\{p_j^s\}$  belonging to the substrate are unimportant here, one
can replace the latter ones by their mean-field values. If the fluid
is in contact with a particle reservoir with a given (excess)
chemical potential $\mu$ and temperature $T$, the partition function
for the liquid is
\begin{eqnarray}
\label{semiinfinite} Z &=& \sum_{\{p^l_j\}}\exp{\left[-\beta
\left({\cal E}-\mu \sum_{j \in {\cal L}} p_j^l\right)\right]} \\
&=& \sum_{\{p^l_j\}}\exp\left[\beta \left( 4\sum_{i\in {\cal S},j\in
{\cal L}}J_{i,j}^{l,s}\rho_{i}^{s}p_{j}^{l} \right. \right. \nonumber \\
 && \left. \left.+{2} \sum_{i,j\in {\cal
L}}J_{i,j}^{l}p_{i}^{l}p_{j}^{l}+\mu\sum_{j \in {\cal L}}
p_j^l\right)\right] \nonumber ,
\end{eqnarray}
where $\rho_i^s\equiv \langle p_i^s\rangle$. Since the solid
substrate  is only weakly influenced by  its surface, we assume
$\langle\rho_{i}^{s}\rangle=\rho_s={\rm const.}$, in  ${\cal S}$ so
that
\begin{eqnarray}
Z&=&\sum_{\{p^l_j\}}\exp\left[\beta \sum_{j\in {\cal L}}
\left(4\rho_s \sum_{i\in {\cal S}}J_{i,j}^{l,s}+\mu \right)
p_{j}^{l}\nonumber \right. \\
&& \left. + {2}\beta \sum_{i,j\in {\cal
L}}J_{i,j}^{l}p_{i}^{l}p_{j}^{l}\right].
\end{eqnarray}
By modeling the pair potentials as
\[
J^l_{i,j}\equiv  J^l/(1+|{\bf r}_i-{\bf r}_j|  ^{d+\sigma})\;
\theta(|{\bf r}_i-{\bf r}_j|  -1),
\]  and
\[J^{l,s}_{i,j}\equiv  J^{l,s}/|{\bf r}_i-{\bf r}_j|^{d+\sigma_s} \; \theta(|{\bf
r}_i-{\bf r}_j|-1)
\]
one finds for $\sum_{i}J_{i,j}^{l,s}$:
\begin{widetext}
\begin{eqnarray}
\sum_{i \in {\cal S}_{1/2}}J_{i,j}^{l,s} &=& J^{l,s}\sum_{i\in {\cal
S}_{1/2}}\frac{1}{|{\bf r}_i-{\bf r}_j|^{d+\sigma_s}}  =
J^{l,s}\sum_{r_1=0}^{\infty}\sum_{r_2=-\infty}^{\infty}
\cdots\sum_{r_d=-\infty}^{\infty}
\frac{1}{\left[(z_j+r_1)^2+r_2^2+r_3^2+\cdots+
r_d^2\right]^{(d+\sigma_s)/2}}\nonumber \\
&\simeq& J^{l,s}\int_0^\infty dr_1\int_{-\infty}^\infty dr_2 \cdots
\int_{-\infty}^\infty dr_d \frac{1}{\left[(z_j+r_1)^2+
r_2^2+r_3^2+\cdots+r_d^2\right]^{(d+\sigma_s)/2}}\nonumber \\
&=&J^{l,s}\pi^{(d-1)/2}
\frac{\Gamma\left(\frac{1+\sigma_s}{2}\right)}
{\sigma_s\Gamma\left(\frac{d+\sigma_s}{2}\right)} z_j^{-\sigma_s},
\label{Jls}
\end{eqnarray}
\end{widetext}
where $z_j \ge 1$ characterizes the distance of the particle $p_j$
from the boundary with the half space ${\cal S}_{1/2}$ occupied by
the substrate. We consider the fluid particles to be in the region
$0 \le z \le L$, where $L$ is the width of the film confined
between the two surfaces. Therefore, the partition function is
\begin{equation}
Z=\sum_{\{p^l\}}\exp\left[\beta \left( \sum_{j\in {\cal
L}}\mu_{j}p_{j}^{l}+{2} \sum_{i,j \in {\cal
L}}J_{i,j}^{l}p_{i}^{l}p_{j}^{l}\right)\right],
\end{equation}
i.e., the system is equivalent to one with a spatially varying
chemical potential $\mu_j=\mu-V_j$ acting on a particle $p_j$ at a
distance $z_j+1$, $0\le z_j \le L$, from the left boundary surface
and at a distance $(L+1-z_j)$ from the right one where $V_j\equiv
V(z_j)$ is given by the superposition
\begin{equation}
\label{Vz} V(z)=  v_s
\left[(z+1)^{-\sigma_s}+(L+1-z)^{-\sigma_s}\right],
\end{equation}
with
\begin{equation}
\label{Avsd} v_s=-4\pi^{(d-1)/2}
\frac{\Gamma\left(\frac{1+\sigma_s}{2}\right)}
{\sigma_s\Gamma\left(\frac{d+\sigma_s}{2}\right)} \rho_s J^{l,s}.
\end{equation}
In the current article we choose such boundary conditions that
$\rho(0)=\rho(L)=1$, where $\rho(z)=\langle p_j^l \rangle$. This is
known as $(+,+)$ boundary conditions. The pressure $p$ in the fluid
as a function of $\{\mu_j\}$ and $T$ follows from $\beta p=|{\cal
L}|^{-1}\ln Z$, where $|{\cal L}|$ is the number of lattice sites in
the region ${\cal L}$. The critical properties of this model can be
directly derived from the known critical behavior of the
corresponding magnetic system that one obtains under the
transformation $m_i=2p_i-1$, where $m_i\in \{-1,1\}$. One arrives at
\begin{equation}\label{pressure}
\beta p = \frac{1}{2}\beta |{\cal L}|^{-1}\left[\sum_{j\in {\cal
L}}\mu_j+\sum_{i,j \in {\cal L}}J_{i,j}^l\right]-\beta f,
\end{equation}
where $f$ is the free energy of the magnetic system
\begin{equation}\label{free_energy}
-\beta f=\ln \sum_{\{m\}}\exp\left[\sum_{j\in {\cal
L}}m_jh_j+\frac{1}{2}\sum_{i,j \in {\cal L}} K_{i,j}\, m_i m_j
\right],
\end{equation}
where
\begin{equation}\label{rel}
K_{i,j}=\beta J_{i,j}^l, \qquad \mbox{and} \qquad h_j=\frac{1}{2}
\beta \mu_j+\sum_{i\in {\cal L}}K_{i,j}.
\end{equation}
The mean-field critical properties of the model (\ref{free_energy})
are well known. The critical exponents are $\beta=\nu=1/2$,
$\gamma=1$ and for the uniform system the critical point is at
$\{h=0,K_c^{-1}=\sum_{i}K_{i,j}\}$. At the critical point
$<m_i>_c=m_c=0$. Thus, for the fluid system we derive that the
corresponding critical point is at $\{K=K_c,\mu=\mu_c\}$, where
$\mu_c=-2\sum_{i\in \mathbb{Z}^d}J_{i,j}^l$ while at this point
$<p_i>_c\,=\rho_c=1/2$. With the help of $\mu_c$ and $\rho_c$ the
corresponding expressions for $h_j$ can be rewritten in the form
\begin{eqnarray}\label{hj}
    h_j &=&\frac{\beta}{2}\left(\mu-\mu_c\right)-
    \frac{\beta}{2}\left(V_j+4\sum_{i \in {\cal S}}J_{i,j}^l
    \rho_c\right) \nonumber \\
    &=& \frac{\beta}{2} \left(\mu-\mu_c\right)+\beta \frac{2
    \pi^{(d-1)/2}}{\sigma \Gamma(\frac{d+\sigma}{2})}
    \left(J^{l,s}\rho_s-J^l \rho_c \right) \nonumber \\
    && \times \left[(z_j+1)^{-\sigma}+
    (L+1-z_j)^{-\sigma}\right].
\end{eqnarray}
From Eq. (\ref{hj}) (see also \cite{DFD}) one identifies that
\begin{equation}
\label{sdef} h_{\rm w,s}=
2\pi^{(d-1)/2}\frac{\Gamma\left(\frac{1+\sigma}{2}\right)}
{\sigma\Gamma\left(\frac{d+\sigma}{2}\right)} \beta(\rho_s
J^{l,s}-\rho_c J^l).
\end{equation}

The equation of the magnetization profile (\ref{eqofstate}) also
directly follows from Eq. (\ref{free_energy}). Denoting
$m_i^*=\,<m_i>$ one obtains
\begin{equation}\label{order_par_profile}
m_i^*={\rm tanh}\left[\sum_j K_{i,j}\; m_j^* +h_j\right].
\end{equation}
Taking into account that $\rho_c=1/2$ one can rewrite $m_i^*$ in
the form $m_i^*=2\rho_i-1=(\rho_i-\rho_c)/\rho_c$ and, thus Eq.
(\ref{order_par_profile}) takes the form
\begin{equation}\label{order_par_profile_rho}
\frac{\rho_i-\rho_c}{\rho_c}={\rm tanh}\left[\sum_j K_{i,j}\;
 \frac{\rho_j-\rho_c}{\rho_c} +h_j\right].
\end{equation}

In what follows we take the ${\rm ^3He}$   or $^4$He atoms to be
constrained by an Au substrate. Then, according to Refs.
\cite{VC81,ZK76,ZK77} $v_s\simeq -270$ meV \AA$^3/r_0^3$, where
$r_0$ is the distance between the helium atom and the Au surface. We
will assume that $r_0$ is the same as the distance between the ${\rm
^3He}$  or $^4$He atoms (but being different for $^3$He and $^4$He
cases, respectively). It is clear that $r_0$ provides the scale of
the length of the unite cell of the lattice on which we consider the
fluid embedded. An estimation of $r_0$ can be obtained from some
general data for ${\rm ^3He}$ or $^4$He. The critical density of
${\rm ^3He}$ is $\rho_c \simeq 0.01375$\;mol/cm$^3\simeq 0.04145$\;g/cm$^3$ \cite{PDM79}, while for $^4$He it is $\rho_c \simeq
0.017399$\;mol/cm$^3 \simeq 0.0690$ \; g/cm$^3$
\cite{LMF2000,R68}, wherefrom one easily derives that at the
critical point one has $8.28\times 10^{27}$ particles/m$^3$ for
$^3$He and $1.38\times 10^{28}$ particles/m$^3$ for $^4$He. This
leads to the conclusion that the space ``allocated" for one particle
is of the order of 120.77\;\AA$^3$ for $^3$He and of the order of
72.55\;\AA$^3$ for $^4$He, i.e. the size of one ${\rm ^3He}$ atom
at the critical point is of the order of 4.9\;\AA while for
$^4$He it is 4.2\;\AA. Thus, for $v_s$ one has $v_s\simeq -2.3$
meV $\simeq - 3.68 \times 10^{-22}$\;J for $^3$He and $v_s\simeq
-3.6$ meV $\simeq - 5.82 \times 10^{-22}$\;J for $^4$He. For $T$
near the critical temperature $T_c=3.3$ K of ${\rm ^3He}$
\cite{PDM79} one has $k_B T_c\simeq 4.55 \times 10^{-23}$ J and thus
$\beta \simeq \beta_c \simeq 2.2 \times 10^{22}$ J$^{-1}$ with $
\beta_c v_s \simeq -8.1$. For $^4$He $T_c=5.2$ K \cite{LMF2000,R68}
and therefore $k_B T_c\simeq 7.17 \times 10^{-23}$ J, $\beta \simeq
\beta_c \simeq 1.4 \times 10^{22}$ J$^{-1}$, and $ \beta_c v_s
\simeq -8.1$. Taking into account that the atomic weight of Au is
$196.97 \;u$, whereas its density is $19.3$ \;g/cm$^3$ and having
in mind that the atomic weight of ${\rm ^3He}$ is 3 $u$ and that the
atomic weight of $^4$He is 4 $u$ (where $u=1.6605 \times 10^{-27}$\;kg
is the atomic mass unit), it is easy to verify that the number
density of Au is $7.1$ times larger than the number density of the
${\rm ^3He}$ and $5.7$  times larger than that of $^4$He at the
critical point of the respective fluid. Since within the mean-field
theory, the number density of both ${\rm ^3He}$ and $^4$He at their
respective bulk critical points is $\rho=\rho_c=1/2$, we obtain that
$\rho_s\simeq 3.55$ for $^3$He films and $\rho_s\simeq 2.85$ for
$^4$He films. As an estimate of $J^{l,s}$ one immediately derives
from Eq. (\ref{Avsd})  (for $d=\sigma_s=3$) the result that that
$J^{l,s}\simeq 4.95 \times 10^{-23}$\;J for $^3$He and that
$J^{l,s}\simeq 9.75 \times 10^{-23}$\;J for $^4$He. Next,
neglecting the contribution due to $J^l$, i.e. the interaction
between the atoms of ${\rm ^3He}$ and also between the atoms of
$^4$He, one finds that $h_{\rm w,s}\simeq -\frac{1}{2}\beta_c v_s
\simeq 4.05$ both for $^3$He and $^4$He films, i.e. the surface
field is, indeed, relatively large and cannot be neglected.

Next, we justify the approximation made for $J^l$.
Within the mean-field approximation we have $\beta_c J^l=0.160$.
Thus, from the experimentally known value of $\beta_c$ we conclude
that $J^l\simeq 7.3 \times 10^{-24}$\;J for $^3$He and $J^l\simeq
1.14 \times 10^{-23}$\;J for $^4$He. Note that these estimates
are very close to those based on the general expectation that
$k_B T_c \sim J^l$, which leads to $J^l \sim 10^{-23}$\;J.
Therefore, $J^{l,s}/J^l\simeq 6.6$ for $^3$He and $J^{l,s}/J^l\simeq
8.5$ for $^4$He, i.e. the interactions of the atoms of ${\rm ^3He}$
and $^4$He with the Au substrate are much stronger than the
interactions between themselves. If, nevertheless, one insists on
taking these interactions into account a simple calculation shows
that $h_{\rm w,s}$ changes from $h_{\rm w,s}=4.05$ to $h_{\rm
w,s}=3.96$ for $^3$He and from $h_{\rm w,s}=4.05$ to $h_{\rm
w,s}=3.97$ for $^4$He. Summarizing, one can conclude that the
surface field has almost the same value for both the $^3$He and
$^4$He films bounded by Au surfaces, and thus one can predict that
the finite-size behavior of their finite-size susceptibilities for a
given fixed $L$ will be practically indistinguishable for both
fluids.

We finish this Appendix by briefly commenting on the correlation
length amplitudes for the correlation lengths $$\xi_t(T)\equiv
\xi_\infty(T\rightarrow T_c^\pm,h=0)\simeq \xi_0^\pm |t|^{-\nu}$$
and $$\xi_h(h)\equiv \xi_\infty(T_c, h\rightarrow 0)\simeq
\xi_{0,h}|h|^{-\nu/\Delta}.$$ One can show that in the case of a van
der Waals fluid-fluid potential the amplitude $\xi_0^+$ of the
second moment correlation length is \cite{DFD}
\begin{equation}\label{xi0}
\xi_0^+ =\left [ \frac{\frac{1}{2d}\sum_{{\bf
r}}\frac{r^2}{1+r^{d+\sigma}}}{\sum_{\bf
r}\frac{1}{1+r^{d+\sigma}}}\right ]^{1/2}.
\end{equation}
Furthermore one has
\begin{equation}\label{xi_other}
\xi_0^-=\xi_0^+/\sqrt{2}, \qquad \mbox{and} \qquad
\xi_{0,h}=\xi_0^+/\sqrt[3]{3}.
\end{equation}
The numerical evaluation of the sum (\ref{xi0}) for $d=\sigma=3$ in
the case of a simple cubic lattice then gives $\xi_0^+=0.635\; a$.
Taking into account that, as we derived above, $a=4.9$\;{\AA} for
${\rm ^3He}$ and $a=4.2$\;{\AA} for $^4$He we obtain that
$\xi_0^+=3.11$\;{\AA} for ${\rm ^3He}$ and that $\xi_0^+=2.67$\;{\AA}
for ${\rm ^4He}$. Of course the procedure used to calculate the
above numbers constitutes a very strong approximation and one shall
not expect to reproduce the best known values of these quantities.
Nevertheless, the comparison with the known data reported in the
literature $\xi_0^+=2.71$\;{\AA} \cite{ZBH2003,GPLE2006} for ${\rm
^3He}$, and $\xi_0^+=2.0$\;{\AA}\cite{KM98} for ${\rm ^4He}$, shows
that the above results are not too bad.  One straightforward way to
improve the above approximation is to consider the fluid imbedded
not on a simple cubic but on a body centered cubic lattice which is
probably ``closer" to the reality since then the atoms are more
closely pact. For such a lattice we obtain $\xi_0^+=0.574\;a$, and
thus $\xi_0^+=2.811$\;{\AA} for ${\rm ^3He}$,
and $\xi_0^+=2.409$\;{\AA} for ${\rm ^4He}$.
These results are indeed essentially close to
the ones obtained by using much more elaborate methods
\cite{ZBH2003,GPLE2006,KM98}.

\end{document}